\documentclass[a4paper,11pt]{article}

\usepackage{jcappub} 

\usepackage{slashed}

\title{The role of self-interacting right-handed neutrinos in galactic structure}

\author[a,b,1]{C.~R.~Arg\"uelles,\note{Corresponding author.}}
\author[c,d]{N.~E.~Mavromatos,}
\author[a,e]{J.~A.~Rueda,}
\author[a,e]{and R.~Ruffini}

\affiliation[a]{International Center for Relativistic Astrophysics Network, P.zza della Repubblica 10, I--65122 Pescara, Italy}
\affiliation[b]{Grupo de Astrof\'isica, Relatividad y Cosmolog\'ia, Facultad de Ciencias Astron\'omicas y Geof\'isicas, Universidad Nacional de La Plata and CONICET, Paseo del Bosque S/N 1900 La Plata, Pcia de Buenos Aires, Argentina}
\affiliation[c]{Theoretical Particle Physics and Cosmology Group, Department of Physics, King's College London, Strand, London WC2R 2LS, UK}
\affiliation[d]{Theoretical Physics Department, CERN, CH 1211 Geneva 23, Switzerland}
\affiliation[e]{Dipartimento di Fisica and ICRA, Sapienza Universit\`a di Roma, P.le Aldo Moro 5, I--00185 Rome, Italy}

\emailAdd{carlos.arguelles@icranet.org}
\emailAdd{Nikolaos.Mavromatos@cern.ch}
\emailAdd{jorge.rueda@icra.it}
\emailAdd{ruffini@icra.it}

\abstract{It has been shown previously that the DM in galactic halos can be explained by a self-gravitating system of massive keV fermions (`inos') in thermodynamic equilibrium, and predicted the existence of a denser quantum core of inos towards the center of galaxies. In this article we show that the inclusion of self-interactions among the inos, modeled within a relativistic mean-field-theory approach, allows the quantum core to become massive and compact enough to explain the dynamics of the S-cluster stars closest to the Milky Way's galactic center. The application of this model to other galaxies such as large elliptical harboring massive central dark objects of $\sim 10^9 M_\odot$ is also investigated. We identify these interacting inos with sterile right-handed neutrinos pertaining to minimal extensions of the Standard Model, and calculate the corresponding total cross-section $\sigma$ within an electroweak-like formalism to be compared with other observationally inferred cross-section estimates. The coincidence of an ino mass range of few tens of keV derived here only from the galactic structure, with the range obtained independently from other astrophysical and cosmological constraints, points towards an important role of the right-handed neutrinos in the cosmic structure.}

\begin{document}
\maketitle
\flushbottom

\section{Introduction}
\label{sec:intro}

The Cold Dark Matter (CDM) model of the Universe, characterized by ordinary matter (about 5\%), a vacuum dark energy (more than 68\%), and Dark Matter (DM, 27\%), with an equation of state resembling a positive-cosmological-constant ($\Lambda$) type fluid (a.k.a. $\Lambda$CDM model) seems to be, at least currently, the cosmological scenario that best fits the plethora of the available cosmological and astrophysical data \cite{2014A&A...571A..16P}. At present, the nature of DM still elude us. Supersymmetry, which provides leading candidates for cold DM, has not been discovered as yet, thus prompting us to consider alternative candidates for DM such as axions, or sterile right-handed neutrinos with masses higher than 100~keV.

\begin{figure}
\centering
\includegraphics[width=0.7\hsize,clip]{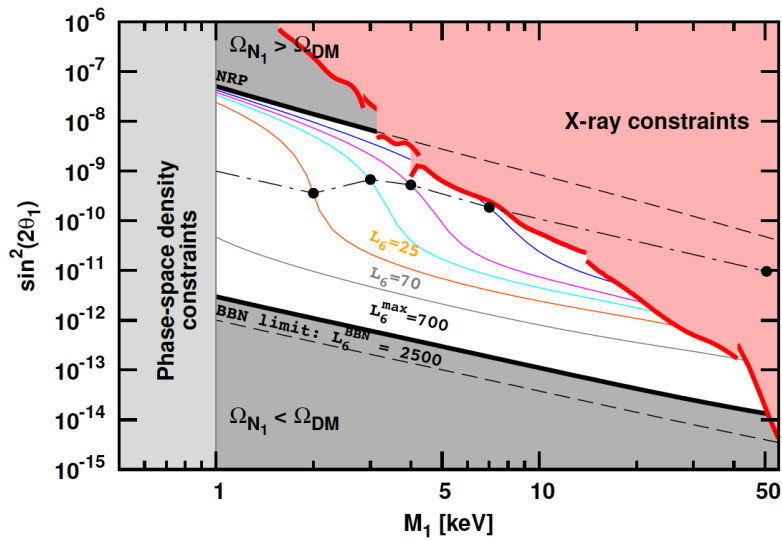}
\caption{Cosmological constraints on the mass ($M_1$) and mixing ($\theta_1$) parameters of the lightest sterile neutrino state $N_1$ of the $\nu$MSM model, consistent with all the current astrophysical and cosmological data~\citep{2005PhLB..631..151A,2009PThPh.122..185S,2009ARNPS..59..191B,2009JCAP...05..012B,2009PhRvL.102t1304B}. This picture was taken from the original version in \cite{2009ARNPS..59..191B}.}
\label{fig:DMsterile}
\end{figure}

On the other hand, right-handed neutrinos with masses less than 50~keV may still play a role in particle physics today, as conjectured in the so-called right-handed neutrino minimal (non-supersymmetric) extension of the standard model ($\nu$MSM) proposed in \cite{2005PhLB..631..151A}, which has been proposed as a viable model for the so-called warm DM (WDM). This model involves three right-handed neutrino states, in addition to the three left-handed active neutrinos of the standard model (SM) sector, of which the lightest, of mass at most a few tens of keV, can live longer than the age of the Universe, thus constituting a viable DM candidate. Such relatively light right-handed neutrinos appear compatible with cosmological DM and Big-Bang-Nucleosynthesis constraints, provided their mixing angles with the active neutrinos of the SM sector are sufficiently small, as shown in figure~\ref{fig:DMsterile}. In general terms, the model appears to be consistent with a plethora of diverse astrophysical and cosmological data~\cite{2009PThPh.122..185S,2009ARNPS..59..191B,2009JCAP...05..012B,2009PhRvL.102t1304B}.

From the astrophysical point of view, the long-pursued study of galactic DM within the context of fundamental physical principles including thermodynamics and statistical physics dates from several decades already (see e.g. \cite{2008gady.book.....B}), given that DM halos show clear universal properties (\cite{2009Natur.461..627G} and refs. therein) and are well fitted by different phenomenological profiles that resemble isothermal equilibrium spheres \cite{2008AJ....136.2648D,2011AJ....142..109C,2014MNRAS.442.2717D,2015ARep...59..656S}. Due to the collisionless nature of DM particles at large scales, it has been recognized that the main mechanisms for  the (quasi) relaxation of a DM halo within observable time-scales are collisionless processes such as phase-mixing and violent relaxation \cite{1967MNRAS.136..101L,2008gady.book.....B}. In contrast to the standard collisional scenarios relevant for (stellar-dominant) globular clusters, violent relaxation takes place within a much shorter timescale, appropriate for the time-varying (global) gravitational potential, leading the system to a long-lived quasi-stationary-state (QSS) that, as shown in \cite{1967MNRAS.136..101L,2002astro.ph.12205C,2002PhRvE..65e6123C,2005A&A...432..117C,2006PhyA..365..102C}, under sufficient mixing conditions can be well described in terms of Fermi-Dirac statistics. More specifically, this kind of Fermi-Dirac (coarse-grained) phase-space distribution can be obtained from a maximization entropy principle at fixed halo mass and temperature as clearly demonstrated in \cite{1998MNRAS.296..569C,1999GReGr..31.1105B,2002astro.ph.12205C,2002PhRvE..65e6123C,2005A&A...432..117C},\footnote{This procedure certainly implies the necessity for these QSS to be bounded in radius. This kind of configurations can be easily obtained within our model, by setting a cutoff in the momentum space of the original Fermi distribution as first realized in \cite{1992A&A...258..223I}, or, more recently and within our relativistic formalism in \cite{2013IJMPD..2260008R,2015mgm..conf.1730F}. In any case the main conclusions presented here do not depend on such a
cutoff that only fixes the outermost halo boundary conditions. Therefore, we shall adopt throughout this paper (for simplicity) the standard Fermi-Dirac statistics, with the momentum cutoff set to infinity.}, leading to mass-density equilibrium distributions typically composed by a condensed core surrounded by a dilute halo. This mechanism was first derived for classical particles (i.e. stars) in \cite{1967MNRAS.136..101L,1978ApJ...225...83S}, and then extended for indistinguishable fermionic particles in \cite{1996ApJ...466L...1K,2002PhRvE..65e6123C}. More recently, and within the context of fermionic DM, it has been argued \cite{2002PrPNP..48..291B,2008PhRvD..77d3518B,2013NewA...22...39D,2013pdmg.conf30204A,2014JKPS...65..801A,2014MNRAS.442.2717D,2015ARep...59..656S,2015MNRAS.451..622R} that a system of self-gravitating fermions, which we have referred to as inos, with masses in the keV regime, plays an important r\^ole in galactic structures. In the more general case of fermionic models allowing for central degeneracy \cite{2013pdmg.conf30204A,2014IJMPD..2342020A,2014JKPS...65..801A,2015ARep...59..656S,2015MNRAS.451..622R}, the density of the inos, which we propose here to be identified with right-handed Majorana neutrinos
\footnote{These neutrinos could be of the DM type appearing in $\nu$MSM~\cite{2005PhLB..631..151A}, but such an identification is not binding. Indeed, our inos can be also identified with sterile neutrinos which do not rely on active-sterile mixing, as the ones analyzed in \cite{2015PhRvD..92j3509P}, and thus consistent with all current cosmological/astrophysical constraints for masses in the keV - MeV range, similarly to the range obtained in section (3.2).}
, shows three physical regimes as a function of the distance from the center: 1) an homogeneous inner core where the inos are in a degenerate state; 2) the uniform density of the core is followed by a steeply decreasing density and an extended plateau in an intermediate region in which the ino's description needs still some quantum corrections; 3) and finally it ends with an asymptotic $\rho\propto r^{-2}$ classical Boltzmann regime. The latter regime is the responsible for the flatness of the rotation curves and therefore it has to fulfill, as an eigenvalue problem, a defined value of the circular velocity. It was further shown in \cite{2015MNRAS.451..622R} that this eigenvalue problem allows to determine the mass of the ino as well as the radius and mass of the inner quantum core. This kind of core-halo structure for DM in galaxies is consistent with the results obtained in \cite{1998MNRAS.296..569C,2002astro.ph.12205C,2002PhRvE..65e6123C} within a pure statistical approach. Interestingly, similar core-halo distributions have been obtained in modern 3-D numerical simulations in the framework of quantum-wave DM approaches \cite{2014PhRvL.113z1302S}. Moreover, such structures appear to characterize more generally long-range collisionless interacting systems, including plasmas and kinetic spin models \cite{2014PhR...535....1L}.

The initial conditions for such a core-halo galactic QSS are provided by the aforementioned collisionless relaxation processes, which imply specific fermionic phase-space distributions as the ones used here. These quantum fermionic distribution functions, with relatively large values of central degeneracy parameters, stabilize the galaxy structures by avoiding a thermodynamic runaway, and thus the gravothermal catastrophe, thanks to the Pauli principle. This is in contrast to what happens in the case of Boltzmann-like configurations~\cite{2014MNRAS.442.2717D}, where a gravothermal catastrophe (similar to one occurring in globular clusters) is an inevitable outcome, even for collisionless DM particles \cite{1990PhR...188..285P,1998MNRAS.296..569C}. The precise choice of the free parameters for the appropriate Fermi distribution, once the QSS is achieved,
is dictated by the correct values of the observed halo parameters (circular and/or dispersion velocities, total mass), as well as the desired quantum core mass and compactness, to account for the observed properties of the central massive object.

The novel approach introduced in \cite{2015MNRAS.451..622R} was applied to different types of galaxies ranging from dwarfs to large spiral galaxies. For ino masses $m\sim10$~keV/$c^2$, one finds excellent agreement with the DM halo observables (see \cite{2015ARep...59..656S,2015MNRAS.451..622R}, for details). At the same time, the approach is capable of providing a theoretical correlation between the inner quantum core and the halo mass, which can be compared with observations \cite{2015MNRAS.451..622R}. We also evaluated the possibility of an alternative interpretation to the black hole in SgrA*, in terms of the high concentration of DM in the inner quantum core. We concluded that, although a compact degenerate core mass $M_c\sim4\times 10^6 M_\odot$ is definitely possible with an ino of $m\sim10$~keV$/c^2$, the core radius is larger by a factor $\sim 10^2$ than the one obtained from the observational limits imposed by the of S-star trajectories such as S1 and S2 orbiting around SgrA* \cite{2008ApJ...689.1044G,2009ApJ...707L.114G}.
To solve this problem, we propose here the inclusion of specific (self) interactions among the inos, which, as we shall demonstrate below,
allows for higher central degeneracies and higher compactness of the inner quantum core. At this point it is important to stress that, already in \cite{1978ApJ...225...83S}, two-particle interactions were predicted to be non-negligible within the dense degenerate cores, due to the appearance of the exclusion principle, in agreement with the ansatz considered here. Moreover, the necessity for considering self-interactions in dense and very-low temperature fermionic systems, such as the ones studied in the current work, has been proven in laboratory experiments. Indeed, in \cite{2008RvMP...80.1215G}, it was argued that the behavior of ultra-cold atomic collisions in (effective) Fermi gases, such as $^6$Li, can be explained in terms of a grand-canonical many-body Hamiltonian with a term accounting for the (spin-enhanced) fermion-fermion interaction. At temperatures of a fraction of the Fermi energy, or, equivalently, for thermal de-Broglie wavelengths larger than the inter-particle mean distances, the self-interactions of the fermions become relevant, in analogy with the situation encountered in our self-interacting neutrino model. However, while in the case of laboratory physics an external trapping potential (such as the one due to magnetic fields) is needed, in the context of DM in galaxies, trapping is ensured by gravity.

The idea of self-interacting DM was first implemented in \cite{2000PhRvL..84.3760S,2001ApJ...547..574D} for cold DM particles with rest masses above $1$~MeV$/c^2$ (and up to $10$~GeV$/c^2$), consistent with the nature of the interactions and the mean free paths considered. In those works, self-interactions were applied uniquely at DM halo scales with typical densities of order $10^{-2} M_\odot/$pc$^3$, suggesting that total cross-sections over the particle mass of order $\sigma/m\sim0.1 - 100$~cm$^2$/g, would imply observational effects in the inner halo regions. It was further shown that a self-interacting DM regime with these values of $\sigma/m$, would generate shallower inner DM profiles, with a consequent reduction in the amount of sub-structures, thereby alleviating important problems of collisionless $\Lambda$CDM simulations, such as the core-cusp~\cite{2010AdAst2010E...5D} and the missing satellite problems~\cite{2011PhRvD..83d3506P}. However, at the same time, some tension with upper limits in the DM cross sections obtained from lensing analysis at galactic cluster scales has emerged. More recently, in \cite{2013MNRAS.430...81R}, motivated by a more refined analysis of the Bullet Cluster~\cite{2008ApJ...679.1173R},  a set of cosmological simulations within CDM were performed,
with the aim of studying further the effects of self-interacting dark matter (SIDM) on density cores of galaxies and galaxy clusters, concluding that $\sigma/m \sim 0.2$~barn~GeV$^{-1}= 0.1$~cm$^2$~g$^{-1}$ is consistent with all the observational constraints.

In the above works, the interactions of DM were modelled by pure classical mechanics descriptions, without making any reference to the details of the interactions. By the contrary, in the present paper, we analyze the possible consequences caused by a \emph{self-interacting relativistic field theoretical model} of Majorana fermions, with vector type interactions and fermion rest-masses in the keV$/c^2$ range, which can play the r\^ole of WDM in galaxies. In particular, we maintain the collisionless nature of the DM fermions at halo scales, and study the two-particle self-interaction effects for different interaction strengths, but only in the (sub-pc) region, where the dense fermionic quantum core arises \cite{2015MNRAS.451..622R}, reaching central densities as large as $10^{16-23}~M_\odot$~pc$^{-3}$. Our method is fundamentally different with respect to the one in other approaches (see, e.g., \cite{2013MNRAS.430...81R}). In such works the ansatz of self-interacting DM is assumed to study its effects on the central parts of halo density profiles coming from cosmological simulations, while we use that ansatz to study the self-interacting effects on the quantum cores of the DM profiles arising from first principle physics such as quantum statistics, thermodynamics and gravitation. At halo-distance scales, our assumption of a collisionless DM is understood by the fact that, the non-interacting fermionic DM distribution, leads naturally to cored inner halos (see figures~\ref{fig:1}--\ref{fig:3} below). Thus, within our fermionic DM model there is no need to make use of the SIDM hypothesis for the halo, as needed in standard $\Lambda$CDM cosmological simulations to alleviate the tension with observations. At the end of the concluding section \ref{sec:out}, we shall comment on how the above mentioned core-cusp discrepancy may be tackled by the current model, and compare briefly our results with predictions made by the standard cosmological N-body simulations of cold and warm DM models. We should stress though that, although many features related to the galactic structure, where the CDM model was challenged, may be explained by our self-interacting fermion (right-hand neutrino) model, the latter should not be viewed as a {\it panacea} for solving all the current open issues regarding that front. There may well be more than one DM species in the universe, and in this work we offer a proposal towards a solution to some important problems in galactic structure within the context of our self interacting (right-handed neutr)-``inos''. 

Both approaches, those based on standard cosmological simulations and ours, can be thought as \textit{complementary} in attacking the problem of the distribution of DM in galaxies. Thus, it is of interest to compare and contrast any theoretical prediction of a self-interacting DM model with the parameters inferred from observations (and/or simulations), such as the total cross-section per unit mass $\sigma/m$ . For this purpose, we consider in this work DM self-interactions mediated by a massive-vector in the dark sector of minimal extensions of the Standard Model (SM), with a $SU(3)$x$SU(2)_{L,R}$x$U(1)$ group invariance, and compute the total cross-section $\sigma$ through an electroweak-like formalism ( see, appendix ~\ref{app:B}, for details). Then, we compare and contrast our theoretical predictions with the observational constraints in section \ref{sec:3.3}. We show that, the requirement that the cross-section per unit mass agree with the one constrained from observations and SIDM cosmological simulations~\cite{2013MNRAS.430...81R}, implies an allowed interaction-strength window $C_V \equiv (g_V/m_V)^2\in (2.6\times10^8,7\times10^8) G_F$ (where $G_F\approx 10^{-5}$~GeV$^{-2}$ is the Fermi coupling constant of the weak interaction), for particle masses in the range $m\in(47,350)$~keV, which as we argue here are in agreement with Galactic-core observables. Here $g_V$ and $m_V$ are the coupling constant of the interaction and the mass of the vector-meson mediator, respectively. Since our approach is not bound to standard $\Lambda$CDM-based conclusions, we also discuss further constraints on the interaction coupling associated with more extreme quantum-core effects. By linking the cross-section with the scattering-rate per particle $\Upsilon$, we also show here an absolute lower bound for the DM interaction strength (and $\sigma$), by calculating the scattering probability among the inos to occur at least once in the age of the galaxy. In this way, we obtain a minimum value for the coupling constant $C_V$.\footnote{It is understood that the coupling $C_V$ ``runs'' with the energy of the SIDM particle, but here we give its value technically at zero momentum. Within our low energy approximations the running of $C_V$ with the energy is very soft and negligible.}

The structure of the article is as follows: in the next section \ref{sec:rhn} we introduce the model of right-handed (Majorana) neutrinos with vector self-interactions, which could be either due to a vector field or describe contact current-current type of interactions. A numerical study of the induced core-halo structure for galaxies, assuming that the above model is the correct one to describe the DM in the Galaxy, is given in section \ref{sec:num}. The effects of the self-interaction in ensuring higher central degeneracies and higher compactness of the inner quantum galactic cores are demonstrated. Finally, discussion of the results and outlook are presented in section \ref{sec:out}. In particular, we specify the order of the vector interactions field strength as well as the minimum value of the fermion masses ($\sim47$~keV/$c^2$) for the model to provide a description of the core-halo structure in a variety of galaxies, from spiral to large elliptical, in agreement with observations. In case the inos are identified with the lightest right-handed Majorana neutrino in the $\nu$MSM, then there is only a narrow (but non trivial) regime of masses for which the model can be consistent with astrophysical/cosmological/galactic data in the sense considered in this paper and in particle physics applications of the $\nu$MSM. Some technical aspects are given in appendices.

\section{Self-interacting right-handed neutrinos} \label{sec:rhn}

We consider in this work a model for SIDM that is a minimal (non supersymmetric) extension of the Standard Model with a sterile neutrino. Such models are reminiscent (but different) of the general idea of the $\nu$MSM~\cite{2005PhLB..631..151A,2009PThPh.122..185S,2009ARNPS..59..191B}. Unlike $\nu$MSM, we allow our right-handed neutrinos to be self-interacting. In particular we concentrate on the lightest of the right-handed Majorana neutrino $N_1$,
which plays the role of DM, and we  introduce phenomenologically, self-neutrino interactions through a massive-vector-meson $V_\mu$ mediator.
Our results regarding the DM particle creation mechanism, are only subject to the assumption that the fermions are of Majorana type (but the formalism is readily extendable to Dirac). This is the common feature we share with the $\nu$MSM, which from our point of view  is another interesting and well-studied case of sterile neutrinos that we use for comparison, given the intriguing similarity of the allowed range of the sterile neutrino DM mass,  $O(10^1)$~keV, which in our case is obtained from a very different approach.

The Lagrangian of the right-handed neutrino sector, including gravity, reads (we use units $\hbar=c=1$):
\begin{equation}
{\mathcal L}={\mathcal L}_{GR}+{\mathcal L}_{N_{R\,1}}+{\mathcal L}_V+{\mathcal L}_{I}\,
\label{eq:Ltotal}
\end{equation}
where
\begin{eqnarray}
{\mathcal L}_{GR} &=& -\frac{R}{16\pi G},\\
{\mathcal L}_{N_{R\,1}} &=& i\,\overline{N}_{R\, 1}\gamma^{\mu}\, \nabla_\mu\,N_{R\,1}-\frac{1}{2}m\,\overline{N^c}_{R\, 1}N_{R\,1},\\
{\mathcal L}_V &=& -\frac{1}{4}V_{\mu\nu}V^{\mu\nu}+\frac{1}{2}m_V^2V_{\mu}V^{\mu} \label{eq:Lterms},\\
{\mathcal L}_{I}&=&-g_V V_\mu J_V^\mu=-g_V V_\mu \overline{N}_{R\, 1}\gamma^{\mu}N_{R\,1}\, ,\label{eq:Lint}
\end{eqnarray}
with $R$ the Ricci scalar for the static spherically symmetric metric background
\begin{equation}\label{metric}
g_{\mu \nu}={\rm diag}(e^{\nu},-e^{\lambda},-r^2,-r^2\sin^2\theta)~,
\end{equation}
where $e^{\nu}$ and $e^{\lambda}$ depend only on the radial coordinate, $r$. The quantity $m$ is the mass of the sterile neutrino, $\nabla_\mu=\partial_\mu\, -\, \frac{i}{8}\, \omega_\mu^ {ab}[\gamma_a, \gamma_b] $ is the gravitational covariant derivative acting on a Majorana spinor, with $\omega_{\mu }^{ab}$ the spin connection and $[ \, , \, ]$ the commutator. The right-handed sterile neutrinos $N_{R\,1}$ satisfy the Majorana four-spinor condition, $\Psi^c=\Psi$, together with $\overline{\Psi}=\Psi^T C$, where the conjugate spinor field $\Psi^c = C \overline{\Psi}^T $ and
$C$ is the unitary ($C^\dagger = C^{-1}$) charge conjugation operator, flipping the fermion chirality, i.e. $(\Psi_L)^c = (\Psi^c)_R $ is right-handed (R), whilst $(\Psi_L)^c = (\Psi^c)_L $ is left-handed (L). The definition of chirality (handedness) is the standard one, $\Psi_{L(R)} = \frac{1}{2} \Big(1 \mp  \gamma^5 \Big) \, \Psi$, with the + (-) sign denoting Right-(Left)handed spinors, and $\gamma_5 = i \gamma^0 \gamma^1 \gamma^2 \gamma^3 $, with $\gamma^\mu$ the $4 \times 4$ Dirac matrices, satisfying $\gamma^\mu \gamma^\nu + \gamma^\mu \gamma^\nu = 2 g^{\mu\nu}$, where $g^{\mu\nu}$ is the (inverse) of the spherically symmetric    space-time metric given above ((\ref{metric})).

The vector-meson mass is $m_V$, whose microscopic origin is not discussed here\footnote{It may well come from an appropriate Higgs mechanism in the dark sector (with a Higgs field that is not necessarily the one of the SM sector).}, and $V_{\mu\nu}=\partial_\mu V_\nu-\partial_\nu V_\mu$, where the ``Lorentz gauge condition'' $\partial^\mu V_\mu = 0$ has been applied for the vector-meson (VM) field $V_\mu$. Notice that the massive-vector-mesons $V_\mu$  should not be viewed as gauge bosons if the fermions are Majorana. As is well known, the Lorentz gauge condition emerges then as a consequence of their equations of motion. Latin indexes denote flat tangent space indexes and are raised and lowered with the Minkowski $\eta_{ab}$ metric.

For simplicity we assume a minimal-coupling form of the vector field with the sterile neutrino current $J^\mu_V$ in the
 interaction term ${\mathcal L}_{I}$ in the lagrangian density. This current is conserved if decays of sterile neutrinos are ignored.
Such a coupling may also arise from linearisation of a Thirring-type four fermion vector current interaction $J^{\mu}_V \, J_{V \mu }$
 by means of an auxiliary vector field $A_\mu$ (which acquires dynamics upon implementing quantum corrections).

In general one may add to (\ref{eq:Ltotal}) a Yukawa term, coupling the (three, in general) right-handed neutrinos to the active neutrino sector (see, e.g., the case of $\nu$MSM~\cite{2005PhLB..631..151A,2009PThPh.122..185S,2009ARNPS..59..191B})
\begin{equation}\label{yuk}
{\mathcal L}_{\rm Yuk} = F_{\alpha I} \, {\overline \ell}_\alpha \, N_{R\, I} \phi^c   + {\rm h.c.}~,  \quad I=1,2,3
\end{equation}
where $\ell_\alpha$ are the lepton doublets of the SM, $\alpha = e, \mu, \tau$, $F_{\alpha I}$ are appropriate Yukawa couplings, and $\phi^c$ is the SM conjugate Higgs field, \emph{i.e}. $\phi^c = i \tau_2 \phi^\star$, with $\tau_2 $ the $2 \times 2$ Pauli matrix. Upon considering such a coupling, one obtains the stringent X-ray and BBN constraints of the mixing angle and mass of $N_{R\, 1}$ depicted in figure~\ref{fig:DMsterile},
given that  (\ref{yuk}) implies decays of the heavy neutrinos $N_I \rightarrow \nu H$, where $H$ denotes the Higgs excitation field, defined via: $\phi = \langle \phi \rangle + H $. In such a case $J^\mu_V$ is \emph{not} conserved in time. However, in the context of $\mu$MSM, the lightest of the heavy neutrinos decay time is longer than the age of the universe, hence the latter can be considered as stable for all practical purposes, thuds playing the r\^ole of dark matter.

For our purposes, as already mentioned, we concentrate here on this lightest neutrino and ignore such a mixing with the SM sector, setting $F_{\alpha 1} = 0$, in which case the lightest neutrino is absolutely stable. The important feature for us are the self-interactions of the right-handed neutrino, which will be used for ensuring phenomenologically correct values for the radius and mass of the galactic core. Since, as we shall see, the mass range we obtain is compatible with the one in figure \ref{fig:DMsterile}, one may switch on the Yukawa term in a full phenomenological study, including the SM sector, and in particular neutrino oscillations and Early Universe physics (e.g.~leptogenesis~\cite{2005PhLB..631..151A,2009PThPh.122..185S,2009ARNPS..59..191B}), without affecting our conclusions. This stems from the very weak nature of the Yukawa couplings $F_{\alpha I}$ as dictated by the seesaw mechanism which is assumed to be in operation here~\cite{2009PhRvL.102t1304B} so as to give a mass in the active neutrinos. For an order-of-magnitude estimate of such $\nu$MSM (subleading) contributions to the effective four-fermion right-handed Majorana neutrino interaction strength we refer the reader to appendix~\ref{app:C}.

A particularly  interesting motivation to include coupling with the SM sector (active) neutrinos $\nu$, is to be able to obtain a possible indirect detection method for the `inos'  through the decaying channel $N_{R\, 1} \rightarrow \nu + \gamma$, with a potential enhancement due to their self-interacting nature~\footnote{In the context of the Yukawa term (\ref{yuk}) such a decay pattern is obtained, e.g., from the decay of the Higgs to two photons.}. Particular attention should be paid  to the recent observations by the Fermi satellite, providing evidence of a clear emission in the energy range $10$--$25$~keV from the central region of the Galaxy \cite{2015PhRvD..92d3503N}. The latter could find plausible explanation by means of a DM particle species with a mass of order $50$~keV/c$^2$, similar to the one obtained here.

Notice that in eq.~(\ref{eq:Lterms}) we included a kinetic term for the VM-field. However, in the mean-field approximation we shall employ in this work, such kinetic terms are irrelevant, thus allowing contact four-fermion interactions among the right-handed neutrinos of Nambu-Jona-Lasinio type to be studied in a similar way. In the latter case, the VM-field is auxiliary.

From (\ref{eq:Ltotal}) one obtains the following equations of motion for the various fields:
\begin{eqnarray}
G_{\mu\nu}+8\pi G T_{\mu\nu} &=& 0\, ,\label{eq:Einstein}\\
\nabla_\mu V^{\mu\nu}+m_V^2 V^\nu-g_V J_V^\nu &=& 0\, ,\\
\overline{N}_{R\, 1}\,i\gamma^{\mu}\overleftarrow{D_\mu}+\frac12 m\overline{N^c}_{R\, 1} &=& 0\, ,
\label{eq:eom1}
\end{eqnarray}
where $G_{\mu\nu}$ is the Einstein tensor and $T_{\mu\nu}$ is the total energy-momentum tensor of the free-fields composed by two terms: $T^{\mu\nu}_{N_{R\,1}}$ and $T^{\mu\nu}_V$, each of which satisfies the perfect fluid prescription
\begin{equation}
T^{\mu\nu}=(\mathcal{E}+\mathcal{P})u^\mu u^\nu-\mathcal{P}g^{\mu\nu}\, ,
\label{eq:EM}
\end{equation}
with $\mathcal{E}$ and $\mathcal{P}$ the energy-density and pressure which we define below.

\subsection{Relativistic mean-field approximation}

We now introduce the relativistic mean-field (RMF) approximation. In this approach, the system can be considered as corresponding to a static uniform matter distribution in its ground state\footnote{As it is shown in section \ref{sec:num}, this approximation is well justified when applied to all the fields (real and mediators) under the physical conditions of the quantum core, which is composed by a very large amount of fermions in a highly degenerate state, in some analogy with the physics of compact objects.}. Thus, the vector-meson field as well as the source currents are replaced by their mean values in this state, which,
on account of space translational invariance, are independent of the spatial coordinates $\vec x$; this and the requirement of rotational invariance imply that no spatial current exists,  and only the temporal component of the current is non zero, i.e. $J_V^\mu \rightarrow \langle J_V^0\rangle=\langle\overline{N}_{R\, 1}\gamma^{0}N_{R\,1}\rangle
= \langle N_{R\, 1}^\dagger N_{R\, 1} \rangle$. The last expression within brackets denotes the finite number density of right-handed neutrino matter times the temporal component of the pertinent (average) velocity.

The RMF approximation allows one to solve the coupled system of differential equations (\ref{eq:Einstein}--\ref{eq:eom1}) rather straightforwardly, to obtain directly the mean-field vector-meson as
\begin{equation}
V_0=\frac{g_V}{m_V^2}J_0^V\,
\label{eq:V0}
\end{equation}
with the notation $\langle V_0\rangle\equiv V_0$ and
\begin{equation}\label{currentvelocity}
\langle J_V^0\rangle\equiv J_0^V=n\,u_0\, ,
\end{equation}
where $u_0=e^{\nu/2}$ is the time-component of the (average) future-directed four velocity vector, and we have used the normalization condition $u^\mu u_\nu=1$.

The Majorana spinors in the RMF approximation can be simply expressed as the corresponding momentum (Fourier) eigen-states with no $x-$dependent terms (see, e.g., \cite{2000csnp.conf.....G}) $\Psi(x)=\Psi(k)\, e^{-ik_{\mu}x^{\mu}}$.
Recalling that we are working here with a system comprising of a very large number $N$ of fermions in thermodynamic equilibrium at finite temperature $T$, we can assume that the fermion number  density is expressed in terms of the Fermi-Dirac distribution function $f(k)$
\begin{equation}
n=e^{-\nu/2}\langle\overline{N}_{R\, 1}(k)\gamma^{0}N_{R\,1}(k)\rangle=\frac{g}{(2\pi)^3}\int d^3k\,f(k)\, .
\label{eq:nNR1}
\end{equation}
where $g$ is a spin-degeneracy factor for the Majorana spinors, the momentum integration is extended over all the momentum space, and $f(k)=(\exp[(\epsilon(k)-\mu)/(k_B T)]+1)^{-1}$. Here $\epsilon(k)=\sqrt{k^2+m^2}-m$ is the particle kinetic energy, $\mu$ is the chemical potential with the particle rest-energy subtracted off, $T$ is the temperature of the heat bath, and $k_B$ is the Boltzmann constant. It is important to notice that we are working with the right-handed component of the full Majorana spinor $\Psi$, and so, although a full Majorana spinor (left plus right chiral states) is its own antiparticle implying a spin degeneracy $g=4$, this is not the case for the singlet right-handed component $N_{R\,1}$ (viewed as a spin $+1/2$ fermion of one helicity state), for which $g=1$. From now on we adopt this value for $g$.

\subsection{Thermodynamic equilibrium conditions and equations of motion}

We now introduce the thermodynamic equilibrium conditions. In the case of a self-gravitating system of semi-degenerate fermions at finite temperature in general relativity, in absence of any self-interactions (other than gravity) such conditions read \cite{2015MNRAS.451..622R}: $e^{\nu/2}T=$constant and $e^{\nu/2}(\mu+m)=$constant. The first equation corresponds to the Tolman condition \cite{1930PhRv...35..904T}, and the second to the Klein condition \cite{1949RvMP...21..531K}. In the presence of the vector-meson mediator interaction (\ref{eq:Lint}), it can be shown that only the Klein condition is modified; the generalized thermodynamic equilibrium conditions are (see, e.g., \cite{2011NuPhA.872..286R}, for details)
\begin{eqnarray}
e^{\nu/2}T &=& {\rm const.}~, \label{eq:Tolman} \\
 e^{\nu/2}(\mu+m)+g_V\,V_0 &= & e^{\nu/2}(\mu+m+C_V\,n)= {\rm const.}
\label{eq:TolmanKlein}
\end{eqnarray}
where the term $g_V\, V_0$ is interpreted as a potential energy associated to the new meson field $V_\mu$. In deriving the middle  equation of (\ref{eq:TolmanKlein}), we have used eqs.~(\ref{eq:V0}) and (\ref{currentvelocity}).

We can then finally write the full system of Einstein equations (\ref{eq:Einstein}) together with the thermodynamic equilibrium conditions (\ref{eq:TolmanKlein}) in the following dimensionless form\footnote{For $C_V=0$, the coupled system of differential equations (\ref{eq:eq1}--\ref{eq:tolman2}) reduces to the standard form presented in \cite{2015MNRAS.451..622R}.}
\begin{align}
\frac{d\hat M}{d\hat r}&=4\pi\hat r^2\mathcal{\hat E}, \label{eq:eq1} \\
\frac{d\nu}{d\hat r}&=2\frac{\hat M+4\pi\mathcal{\hat P}\hat r^3}{\hat r^2(1-2\hat M/\hat r)}, \\	
\frac{d\theta}{d\hat r}&=-\frac{1}{2\beta}\frac{d\nu}{d\hat r}\frac{\left(1+\frac{C_Vm^2}{4\pi^3}\hat n
    -\frac{C_Vm^2}{4\pi^3}\beta\frac{d\hat n}{d\beta}\right)}{\left(1+\frac{C_Vm^2}{4\pi^3}\frac{1}{\beta}\frac{d\hat n}{d\theta}\right)}\, ,\label{eq:eq3}\\
\beta&=\beta_0 e^{\frac{\nu_0-\nu(r)}{2}}\, , \label{eq:tolman2}
\end{align}
where the following dimensionless quantities were introduced: $\hat r=r/\chi$, $\hat n=Gm\chi^2$, $\hat M=G M/\chi$, $\mathcal{\hat E}=G \chi^2 \mathcal{E}$, $\mathcal{\hat P}=G \chi^2 \mathcal{P}$, with $m_p=\sqrt{1/G}$ the Planck mass, and we have introduced the dimensional factor $\chi=2\pi^{3/2}(1/m)(m_p/m)$ with units of length, scaling as $m^{-2}$. We have also introduced the temperature and degeneracy parameters $\beta=k_B T/m$, and $\theta=\mu/(k_B T)$, respectively; we have evaluated the constants of the equilibrium conditions of Tolman and Klein at the center $r=0$, which we indicate with a subscript `0'. We have also introduced the parameter
\begin{equation}\label{cv}
C_V\equiv g_V^2/m_V^2,
\end{equation}
which encodes information about the strength of the coupling of the effective interactions of the fermions (`inos') and the mass of the vector-meson mediator. The total energy-density and pressure $\mathcal{E}$ and $\mathcal{P}$ contained in (\ref{eq:EM}), can be split  in two components,
\begin{equation}
    \mathcal{E}= \mathcal{E}_\mathcal{C}+\mathcal{E}_V\, ,\quad \mathcal{P}= \mathcal{P}_\mathcal{C}+\mathcal{P}_V\, ,\label{eq:Pt}
\end{equation}
with $\mathcal{E}_\mathcal{C}$ and $\mathcal{P}_\mathcal{C}$ the contributions to the energy-density and pressure from fermions in the RMF approximation, calculated as $\langle T_0^0\rangle_{N_{R\,1}}=\mathcal{E}_\mathcal{C}$ and $\langle T_1^1\rangle_{N_{R\,1}}=\mathcal{P}_\mathcal{C}$ respectively. They are fully determined by the distribution function $f(k)$ (with particle helicity $g=1$)\footnote{Alternatively, this contribution to the energy can be expressed as the expectation value of the energy $\langle \overline{\Psi}\gamma_0k_0\Psi \rangle$, where $E(k)\equiv k_0$ are the energy eigenvalues of the corresponding Majorana Hamiltonian (see, e.g., \cite{2000csnp.conf.....G}).}
\begin{align}
&\mathcal{E}_\mathcal{C} = m\frac{1}{(2\pi)^3}\int f(k)\left[1+\frac{\epsilon(k)}{m}\right]\,d^3k,\label{eq:E}\\
&\mathcal{P}_\mathcal{C} = \frac13 \frac{1}{(2\pi)^3}\int
   f(k)\left[1+\frac{\epsilon(k)}{2 m}\right]\epsilon\,d^3k,\label{eq:P}
\end{align}
while
\begin{equation}
\mathcal{E}_V=\mathcal{P}_V=\frac12 e^{-\nu}m_V^2V_0^2 = \frac12C_V\,n^2\, ,
\label{eq:EandPV}
\end{equation}
is the contribution from the VM-field. We shall next proceed to solve the system of equations (\ref{eq:eq1}--\ref{eq:tolman2}), including a discussion on the boundary conditions appropriate for the description of the Milky Way, as a self-consistency check of the approach.

\section{Numerical solutions}\label{sec:num}

We now apply the theoretical formalism presented above to study the DM distribution on different astrophysical objects from spiral to large elliptical galaxies, for given boundary conditions in agreement with observations. At the end, we give the DM particle mass and total cross section constraints arising from the numerical analysis of the boundary-value problem.

\subsection{Spiral galaxies: the Milky Way}

\begin{figure}[!hbtp]
\centering
\includegraphics[width=0.7\textwidth]{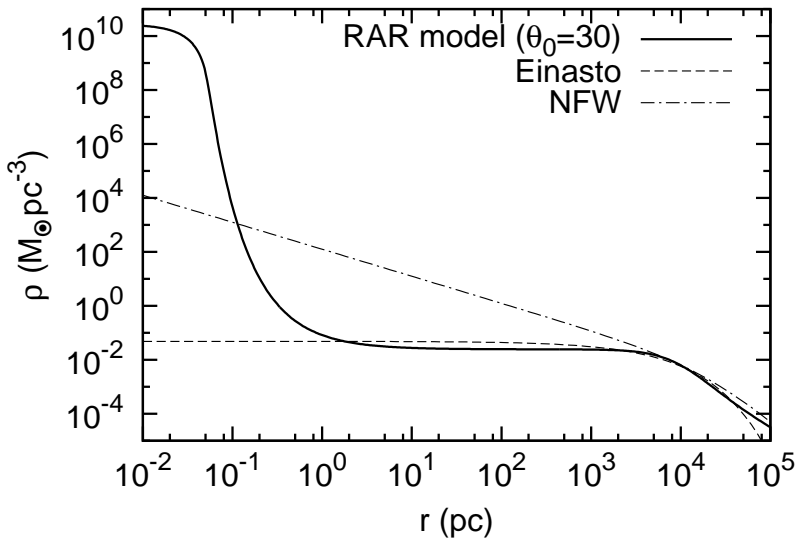}
\caption{Comparison of the Ruffini-Arg\"uelles-Rueda (RAR) DM density profile \cite{2015MNRAS.451..622R} (non-interacting case $C_V=0$) with the Navarro-Frenk-White (NFW) one \cite{1997ApJ...490..493N}, and with a cored Einasto profile \cite{1965TrAlm...5...87E,1989A&A...223...89E}. The RAR profile is here given for the parameters: $\beta_0=1.251\times10^{-7}$, $\theta_0=30$ and $m=10.54$~keV$/c^2$. The NFW profile is $\rho_{NFW}(r)=\rho_0 r_0/[r(1+r/r_0)^2]$ with $\rho_0=5\times10^{-3} M_\odot$~pc$^{-3}$ and $r_0=25$~kpc, and the Einasto profile is given by $\rho_E(r)=\rho_{-2}\exp{[-2n(r/r_{-2})^{1/n}-1]}$,  with $\rho_{-2}=2.4\times10^{-3} M_\odot$~pc$^{-3}$, $r_{-2}=16.8$~kpc, and $n=3/2$. The free parameters have been chosen to describe the typical properties of spiral galaxies \cite{2008AJ....136.2648D,2011AJ....142..109C}. This picture was taken from the original version in \cite{2015MNRAS.451..622R}.}\label{fig:1}
\end{figure}

The boundary conditions in this case are given by the request of the observational agreement of the inner quantum core and halo part with the following Milky Way properties: 1) the compactness of its `dark' center (SgrA*), i.e. massive and compact enough to explain the dynamics of the S-cluster stars closest to the Milky Way's galactic center, 2) the DM outer halo mass $M_h$ and radius $r_h$, and 3) the onset of flat galactic rotation curve with the specific value of the circular velocity $v_h$ at $r_h$. It is important to recall that we define the radius of the inner quantum core $r_c$ as the distance at which the rotation curve reaches its first maximum, and the outer halo radius $r_h$ at the onset of the flattening rotation curve, which occurs at the second maximum (see also figure~1 in Ref.~\cite{2015MNRAS.451..622R}). Notice that the so called \textit{halo radius} (and mass) represent the one-halo scale length (and mass) associated with the fermionic model here presented, and corresponding with the turn-over of the density profiles in total analogy as other halo-scale lengths used in the literature such as $r_0$ or $r_{-2}$ as shown in figure~\ref{fig:1}. The rotation curve is given by the circular velocity
\begin{equation}
v(r)=\sqrt{\frac{G M(r)}{r-2 G M(r)}}.
\end{equation}

Following the above procedure, we shall constrain the physical conditions $\beta_0$ and $\theta_0$, together with the physical parameters, such as the sterile neutrino mass $m$, as well as the coupling parameter $C_V$. We recall that the non-interacting case $C_V=0$ of the model (\ref{eq:Ltotal}) has been recently solved in \cite{2015MNRAS.451..622R,2014JKPS...65..801A}, whose more general DM density profile shows the typical core-halo distribution composed of three different physical regimes as described in the introduction of the present article and demonstrated in figure~\ref{fig:1}.

Indeed, in figure~\ref{fig:1} a solution with $m\sim10$~keV$/c^2$ of the $C_V=0$ non-interacting model of \cite{2015MNRAS.451..622R}, is compared and contrasted with selected DM halo profiles used in the literature. In the sub-parsec core region and for core masses of $\sim 10^6 M_\odot$ typical of (Milky Way-like) galaxies, for an ino mass $\sim 10$~keV$/c^2$, the thermal de-Broglie wavelength, $\lambda_B=h/\sqrt{2\pi m k_B T}$, is larger than the inter-particle mean distance $l$ of the inos, justifying the quantum-statistical nature of the core. A degenerate core with a very low temperature was found in \cite{2015MNRAS.451..622R} to be compatible with the outer halo properties such as the halo radius, mass, and rotation curves of order $10^2~$km/s. In the Boltzmannian region, we have $\lambda_B/l\ll 1$ and, as shown in \cite{2015MNRAS.451..622R}, the specific value of the corresponding circular velocity in the flat region is intimately related to the temperature parameter, $\beta$.

We shall adopt here the ansatz that the self-interactions occur only in the quantum regime and thus within the core, where the thermal de-Broglie wavelength,
\begin{equation}
\lambda_B=\frac{h}{\sqrt{2\pi m k_B T}},
\end{equation}
is larger than the inter-particle mean distance $l$, i.e. $\lambda_B/l>1$. The reader should recall that at this quantum regime, two-particle interactions were predicted to be non-negligible~\cite{1978ApJ...225...83S}. To this end, we set:
\begin{equation}\label{eq:C}
C_V(r) =
\begin{cases}
C_0 &  \quad {\rm at}\quad r<r_m\quad {\rm when}\quad \lambda_B/l>1\, , \\
0   &  \quad {\rm at}\quad r\geqslant r_m\quad {\rm when}\quad \lambda_B/l<1\, ,
\end{cases}
\end{equation}
where $C_0$ is a positive constant and $r_m=r_c+\delta r$ is the core-halo matching point, with $r_c$ the core radius and $\delta r$ the thickness of the core-halo intermediate layer. As we shall show, $\delta r\ll r_c$, and thus the core-halo matching satisfies $r_m\approx r_c$. In the regime $r\geqslant r_m$, where the DM distribution is in a much more dilute state (i.e.~$\lambda_B/l\ll1$), there is the transition from the quantum degenerate state to the Boltzmannian one.

As we show below, the density profile obtained for the interacting case has a similar behavior, with the aforementioned three different regions, as the non-interactive case $C_V=0$ \cite{2015MNRAS.451..622R}. We normalize hereafter the coupling constant $C_0$, for the sake of reference, to the Fermi constant of the SM weak-interaction, i.e. we introduce the dimensionless constant $\overline{C}_0=C_0/G_F$. We define the SM Fermi constant only for normalization purposes, thus $C_0$ must not be thought as a fundamental interaction strength (i.e weak) of the SM. Indeed, the fact that the effective interactions considered here are mediated by a chargeless VM field playing the role of neutral-current interactions through the scattering channel, implies that the inos remain unaffected except for momentum transfer. Therefore, we here adopt a complete phenomenological analysis by studying the maximum possible range of effective interactions strengths which are in agreement with the Milky Way observables.

The mass $M_c$ of the degenerate quantum core must agree with the mass enclosed within the region bounded by the pericenter of the S2 star. At the same time, we use the pericenter of S2 as an upper limit to the core radius $r_{c(S2)}$, i.e. \cite{2008ApJ...689.1044G,2009ApJ...707L.114G}
\begin{equation}
M_c=4.4\times10^6\,M_\odot\, , \quad r_{c(S2)}=6\times10^{-4}\,{\rm pc}\, .
\label{eq:coreObs}
\end{equation}
There is an error of 8\% in the above value of $M_c$ due to the uncertainties in the measurement of the distance to the galactic center $R_0=8.33\pm 0.35$~kpc, while the error in the pericenter of the S2 star is of about 4\% \cite{2009ApJ...707L.114G}. The above parameters imply a central density of order $\sim10^{16} M_\odot/$pc$^3$, which is almost five orders of magnitude larger than the one obtained for the model without self-interactions \cite{2015MNRAS.451..622R} with the same core mass. It is important to make clear that any core radius $r_{\rm Sch}\lesssim r_c\lesssim r_{c(S2)}$ is accepted within our phenomenological treatment, implying central densities in the range $10^{16} M_\odot/$pc$^3\lesssim\rho_0\lesssim 10^{23} M_\odot/$pc$^3$, with $r_{\rm Sch}$ the Schwarschild radius of a black hole of $4.4\times10^6 M_\odot$. Indeed, as we show below, already for an ino mass $m\approx350$~keV/c$^2$ it is possible to obtain a critical core DM core of fully degenerate inos of mass $M_c=4.4\times10^6 M_\odot$ with a radius $r_c\approx 2.5 r_{\rm Sch}$. The critical objects are the last equilibrium configurations, just before undergoing gravitational collapse (see also Ref.~\cite{2014JKPS...65..809A}).

For the observables in the halo region we adopt the fitting procedure outlined in ref.~\cite{2009PASJ...61..153S}. According to that work, the DM best-fit distribution for the Milky Way is provided by the two-parameter Burkert profile with a specific central density parameter $\rho_B^0=2\times10^{-2} M_\odot/$pc$^3$, and a dark halo length scale parameter $h=10$~kpc. For our fermionic model, this corresponds  to a halo radius $r_h$, defined at the maximum of the corresponding rotation curve at the onset of the flat behaviour, and leads to an associated halo velocity $v_h$ and mass $M_h$ given by (the reader is invited to observe the excellent matching between Burkert and fermionic profiles around $r_h$ in  figure~\ref{fig:2}):
\begin{equation}\label{eq:haloObs}
r_h=32.4\, {\rm kpc}\, , v_h=155\, {\rm km/s}\, , M_h=1.75\times10^{11} M_\odot\, ,
\end{equation}
where the subscript $h$ indicates quantities at the halo radius. All the halo parameters are subject to an error of  $\sim10\%$~\cite{2009PASJ...61..153S}. The above value of the circular velocity determinates the value of the temperature parameter at the halo, $\beta_0^h$. For these parameters, we obtain $\beta_0^h=1.065\times10^{-7}$.

We discuss now the core-halo transition. There, the generalized Tolman and Klein equilibrium conditions have to be fulfilled. The Tolman's condition together with the condition imposed by the continuity of the spacetime metric, lead to the continuity of the temperature parameter $\beta(r_m)=\beta_0^h$. Now, from the Klein's condition we can obtain the jump in the degeneracy parameter at the matching point $r_m$, where the (diluted) halo region begins:
\begin{eqnarray}
&\theta(r_m)=\theta_0^h - \frac{C_0 n(r_m)}{m\beta_0^h}\,,\\
&n(r_m)=\frac{\sqrt{2}m^3(\beta_0^h)^{3/2}}{\pi^2}(F_{1/2}+\beta_0^h F_{3/2})\, ,
\end{eqnarray}
where $\theta_0^h$ is the value of $\theta$ from the halo side, and the generalized Fermi-Dirac integrals are evaluated at $r_m$: $F_j=\int_0^\infty dx \, x^j\, (1+\beta_0^h x/2)^{1/2}/[1+e^{x-\theta(r_m)}]$.

\subsection{Other galactic structures: large elliptical galaxies}

In analogy with the Milky Way case, we now apply our SIDM model to other galactic structures, such as large spiral and elliptical galaxies, where clear evidence exists for massive BH-like structures at their centre, together with a DM halo counterpart.



In order to give a \textit{universal} explanation for the galactic DM in terms of an ino mass and an interaction constant (or cross-section), we next apply our theory to larger galaxies for $m=47$~keV, and give the possible values of $C_V$ (\ref{cv}) in agreement with the different galaxy observables. We proceed with two different kinds of typical elliptical galaxies each harboring a different characteristic dark massive object at the center:
\begin{eqnarray}
\centering
& (i) & \qquad M_c=2.3\times10^8 M_\odot\, \qquad \rm{Elliptical} \label{eq:coreObsE},\nonumber\\
& (ii)  &\qquad M_c=1.8\times10^9 M_\odot\, \qquad \rm{Large\,\, Elliptical},\label{example}
\end{eqnarray}
both contained within sub-pc scales \cite{2013degn.book.....M}. Notice that the above cases are representative examples and other intermediate cases between normal spiral galaxies and the large elliptical ones are also contained among the possible solutions of our model (see section \ref{sec:3.2}), but are not given explicitly here for the sake of brevity.~\footnote{Regarding the applicability of our approach to dwarf galaxies, we remark that the rather low central degeneracy values in the phase-space distribution of these systems~\cite{2015MNRAS.451..622R} lead to not very massive cores, therefore not exhibiting massive BH-like features (i.e. with masses $M_c\lesssim 10^4 M_\odot$ for $m\sim 10^1$~keV). Hence, they do not seem to constitute interesting cases for a detailed study of the effects of self-interacting DM.}

The halo parameters are chosen from the observationally inferred correlation between central mass concentrations and dark halo masses ($M_c$-$M_h$) as obtained in \cite{2002ApJ...578...90F} (see figure 5 there). This is in analogy with the study of \cite{2015MNRAS.451..622R}, but in our SIDM case we do reach the upper end of the correlation. Thus, typically, we should have
\begin{eqnarray}
\centering
& (i) &\qquad M_h=4.1\times10^{12} M_\odot\,, \qquad r_h=60\, {\rm kpc}\, ,\qquad \rm{Elliptical},\\
& (ii) & \qquad M_h=1.1\times10^{13} M_\odot\,, \qquad r_h=85\, {\rm kpc}\, \qquad \rm{Large\,\, Elliptical}.
\label{eq:haloObsE}
\end{eqnarray}
The values given above were considered at the one-halo scale-length of our model $r_h$ (located at the maximum of the halo rotation curve), which is similar to the NFW halo scale-length $r_s$ ($r_h\approx r_s/0.6$), where the halo mass was originally obtained in \cite{2002ApJ...578...90F}. These halo magnitudes, combined with
the DM halo morphology of our model, imply  typical halo velocities of $v_h=540$~km/s in case (i), and $v_h=730$~km/s in case (ii). With these values, we finally obtain (analogously as done for the Milky Way in appendix~\ref{app:A}) the following temperature parameters at the halo: $\beta_0^h=1.3\times10^{-6}$, $\beta_0^h=2.4\times10^{-6}$ for (i) and (ii) respectively. The parameters of the model to be used for solving the above boundary conditions, are as follows: we keep the ansatz (\ref{eq:C}) for the interaction constant, while we set $m=47$~keV.

\subsection{Novel DM mass constraints}\label{sec:3.2}

Following the above procedure, we summarize in table~\ref{table:5.2} the solution of the boundary-value problem which fulfills the core and halo observables from Milky Way (\ref{eq:coreObs}),(\ref{eq:haloObs}), and Elliptical and large elliptical galaxies (\ref{eq:coreObsE})--(\ref{eq:haloObsE}) respectively. The calculations were done for the maximum allowed possible range of the interaction constant $\overline{C}_0$, central degeneracy $\theta_0$ and ino mass $m$. Even if the upper limit in the sterile neutrino mass ($m\lesssim 50$~keV$/c^2$) is imposed by cosmological and astrophysical constraints under the assumption of mixing with the SM sector (\emph{cf}. figure~\ref{fig:DMsterile}), we also explore larger (phenomenologically) values of the ino mass, which is possible for sterile neutrinos that do not interact
or have negilgible interactions through a Higgs portal
with the \textit{active} sector.

\begin{table*}
\centering
\begin{tabular}{@{}c|c|c|c|c|c|c@{}}
\hline
\multicolumn{7}{c} {Milky Way ($M_c=4.4\times10^6 M_\odot$)} \\
\hline
$m$ (keV) & $\overline{C}_0$ & $\theta_0$ & $\beta_0$ & $r_c$ (pc) & $\delta r$ (pc) & $\theta(r_m)$\\
\hline

47 & 2 & $3.70\times10^3$ & $1.065\times10^{-7}$ & $6.2\times10^{-4}$ & $2.1\times10^{-4}$ & -29.3 \\
      & $10^{14}$ & $3.63\times10^3$ & $1.065\times10^{-7}$ & $6.2\times10^{-4}$ & $2.2\times10^{-4}$ & -29.3 \\
      & $10^{16}$ & $2.8\times10^3$ & $1.065\times10^{-7}$ & $6.3\times10^{-4}$ & $2.4\times10^{-4}$ & -29.3 \\
\hline
350 & 1 & $2.40\times10^6\,^{(\dag)}$ & $1.431\times10^{-7}$ & $1.3\times10^{-6}$ & $6.7\times10^{-7}$ & -37.3 \\
      & $10^{14}$ & $1.27\times10^5$ & $1.104\times10^{-7}$ & $5.9\times10^{-6}$ & $9.4\times10^{-7}$ & -37.3 \\
      & $4.5\times10^{18}$ & $1.7\times10^1$ & $1.065\times10^{-7}$ & $5.9\times10^{-4}$ & $2.0\times10^{-4}$ & -37.3\\
\hline
\multicolumn{7}{c} {Elliptical ($M_c^{cr}=2.3\times10^8 M_\odot$)} \\
\hline
47  & 2 & $1.76\times10^5\,^{(\dag)}$ & $1.7\times10^{-6}$ & $7.9\times10^{-5}$ & $3.9\times10^{-5}$ & -31.8 \\
    & $10^{14}$ & $5.8\times10^4$ & $1.4\times10^{-6}$ & $1.4\times10^{-4}$ & $4.8\times10^{-5}$ & -31.8 \\
    & $10^{16}$ & $1.5\times10^4$ & $1.3\times10^{-6}$ & $3.0\times10^{-4}$ & $7.0\times10^{-5}$ & -31.8 \\
\hline
\multicolumn{7}{c} {Large Elliptical ($M_c=1.8\times10^9 M_\odot$)} \\
\hline
47 & $10^{16}$ & $1.02\times10^4$ & $3.0\times10^{-6}$ & $3.8\times10^{-4}$ & $1.8\times10^{-5}$ & -32.8 \\
\hline
\end{tabular}
\caption{Set of model parameters for three different galaxy types analyzed that solve the corresponding boundary-value problem imposed by the given galaxy observables, from spiral (Milky Way) to large elliptical galaxies.$^{(\dag)}$ Critical central degeneracy parameters ($\theta_0^{cr}$), associated with the \textit{turning point} or last stable equilibrium solution \cite{2014JKPS...65..809A}.}
\label{table:5.2}
\end{table*}

Two important conclusions can be drawn from the numerical analysis presented in Table~\ref{table:5.2}:

I) For $m<47$~keV$/c^2$ and $m>350$~keV$/c^2$ there is no pair of parameters ($\overline{C}_0,\theta_0$) able to be in agreement with the Milky Way observables. While $m=47$~keV$/c^2$ is the lowest admissible particle mass up to which the core observational constraints are fulfilled (within observational errors), $m=350$~keV$/c^2$ is the uppermost bound set by the reaching of the critical core mass for gravitational collapse \cite{2014JKPS...65..809A}, $M_c^{cr}\propto M_{pl}^3/m^2\approx 4.4\times10^6 M_\odot$, where $M_{pl}$ is the Planck mass. For $m=47$~keV, one reaches the critical mass $M_c^{cr}\sim10^8 M_\odot$ corresponding to a massive BH alternative for elliptical galaxies (when $\mathcal{P}_V<\mathcal{P}_\mathcal{C}$); but it is also possible to reach core mass values as large as $M_c\sim 10^9 M_\odot$ (when $\mathcal{P}_V\sim\mathcal{P}_\mathcal{C}$), and applicable to massive BH alternatives in large elliptical galaxies. In the latter case the quantum core can reach radii as small as $r_c\approx 2 r_{\rm Sch}$ (see last line in table~\ref{table:5.2}), developing interior sound-wave speeds as large as $5 \%$ of the speed of light.

It is worth to notice that for all the galaxy types analyzed, there is a \textit{common window} of interaction coupling constant parameters ($\overline{C}_0$), the value $\overline{C}_0=10^{16}$ constitutes a very interesting case, allowing for a successful \textit{universal} application of the model all the way from spiral to large elliptical galaxies.\footnote{If, instead, the interaction constant is forced to agree with the N-body simulation results for the total DM cross-section (i.e. $\overline{C}_0=7\times10^{8}$ for $m=47$~keV, as detailed in section \ref{sec:3.3}), then the applicability of our model is reduced up to elliptical galaxies with dark compact cores of $\sim2\times10^8 M_\odot$.}

II) As the value of the coupling constant $\overline{C}_0$ increases from unity, the contribution to the total energy and pressure from the meson-vector field ($\sim C_0 n^2$) becomes more and more relevant. For instance, as can be seen in table \ref{table:5.2} in the Milky Way case, for $\overline{C}_0\sim10^{14}$ and for $m=47$~keV, a slightly lower value for the central degeneracy is needed to have the same core mass as compared with the $\overline{C}_0\sim 1$ regime. In other words, if the same central degeneracy as in the former $\overline{C}_0\sim1$ case is used, an increase of $\sim$~few $\%$ in the core mass $M_c$ would appear. For this lower ino mass bound, the self-interactions cannot exceed $\overline{C}_0\sim10^{16}$, because otherwise the now lower central degeneracy needed to compensate for the core mass, would be too low to fulfill with the upper core radius constraint $r_{c(S2)}$. More evident is the case when the ino mass reaches $m=350$~keV$/c^2$, where the highest interaction regime $\overline{C}_0\sim10^{18}$ fulfilling the core radius and mass, is reached at a central degeneracy about two orders of magnitude lower with respect to the $\overline{C}_0=1$ case.


\subsection{Cross-section constraints}\label{sec:3.3}

It is possible to establish, within our theoretical approach, a direct link between the total cross-section $\sigma$ and the interaction strength $C_0=(g_V/m_V)^2$. This will allow us to compare our results  with the ones given in the literature, regarding
the total cross-section per DM mass, $\sigma/m$.  To this end, we shall consider a four-fermion (elastic) scattering for the inos, with a massive-vector boson $V^\mu$ as mediator field.

The total cross-section in the center-of-mass (CoM) system for two incident right-handed Majorana neutrinos $N_{R\,1}$ with four-momentum ($p_i=(E,\bf{p_i})$), $i=1,2$, (with $E^2=p^2+m^2$) which collide through an elastic scattering picture, and produce two final identical particles with momentum ($p'_i=(E',\bf{p'_i=p_i})$), $i=1,2$ ($E'=E$) is given by (see appendix~\ref{app:B} for details)
\begin{equation}
\sigma^{tot}_{CoM}=\left(\frac{g_V}{m_V}\right)^4\frac{1}{4^3\pi^2}[29m^2+89p^2+89/3\frac{p^4}{p^2+m^2}]\, .
\label{eq:totalsigma}
\end{equation}
where we have used $\theta_{W}'=0$ since we only have one massive-vector mediator.

We now calculate the total $N_{R\,1}$-$N_{R\,1}$ scattering cross-section in the quantum core of the Galaxy. For this we use the following approximations leading to a simplified version of eqn.~(\ref{eq:totalsigma}). In a typical quantum-core (non-relativistic) one-particle momentum $p$ is given the Fermi momentum (at the core of the configuration) $p\sim p_F=(3\pi^2\hbar^3\rho_c/m)^{1/3}$. For typical core densities used in this work, $\rho_c\sim10^{16-23} M_\odot/pc^3$, and $m=47$--$350$ keV, which leads directly to the following `low-energy limit' $p^2\ll m^2$ for our particles. This is the opposite with what one generally finds in laboratory collision-particle experiments (i.e a `high-energy limit'), which is easily understood because the sterile neutrinos are in a very low temperature, and a high degenerate regime in the core. With all this, equation (\ref{eq:19}) reads
\begin{equation}
\sigma^{tot}_{core}\approx\frac{(g_V/m_V)^4}{4^3\pi}29m^2 \qquad (p^2/m^2\ll 1)\, .
\label{eq:20}
\end{equation}
Equation (\ref{eq:20}) links the (dimensionless) interaction constant of our inos (expressed relative to the (weak interactions) Fermi constant $G_F=1.166\times10^{-5}$~GeV$^{-2}$), (cf. (\ref{cv})),
\begin{equation}\label{cv2}
\overline{C}_V=\left(\frac{g_V}{m_V}\right)^2 G_F^{-1},
\end{equation}
with the total cross-section and the particle mass. Thus, if we constrain the total cross-section to the N-body simulation value $\sigma^{tot}/m=0.1$~cm$^2$/g \cite{2013MNRAS.430...81R}, our coupling constant $\overline{C}_V$ would be constrained to the value
\begin{eqnarray}
\overline{C}_V\in (2.6\times10^8,7\times10^8),
\end{eqnarray}
for ino masses in the range $m\in(47,350)$~keV. It worths noticing that for $C_V\sim 10^8 G_F$, the mass of the massive-vector meson would be constrained to values $m_V\lesssim 3\times 10^4$~keV, in order to satisfy $g_V\lesssim 1$ as requested by the self-consistency of the perturbation scheme we have applied to compute the cross-section.

We can further try to get an absolute lower bound for $C_V$. Interestingly, it can be obtained by answering the question as to which physical conditions need to be fulfilled by the keV particles in the galaxy for a self-interacting DM regime to appear. A conservative answer one might give is that $\sigma$ should be large enough so that a scattering probability among the inos should occur at least once during the age of the galaxy ($t_{age}$), that is, the product of the scattering-rate per particle ($\Upsilon$) times $t_{age}$ be larger than unity: $\Upsilon t_{age}\gtrsim 1$. For this we first calculate $\Upsilon$, which is linked to the cross-section $\sigma$ via \cite{1990eaun.book.....K}
\begin{equation}
\Upsilon=\sigma |v_{rel}| n,
\label{eq:scattrate}
\end{equation}
where $v_{rel}$ is the relative velocity of the interacting inos and $n$ the particle number density. The above formula can be written as an \textit{order of magnitude} expression as follows: $\Upsilon\sim \sigma/m^2 \,p\, \rho_0 $, with $p$ a typical momentum of the inos in the quantum core (i.e. $p\equiv p_F$) and $\rho_0$ the central density (valid in the low energy regime approximation). Then, by assuming typically $t_{age}\sim 10^{16}$~s (i.e. redshift $z\sim10$ at galaxy formation epochs), one obtains, for $m=47$~keV and $\rho_0\sim10^{16} M_\odot$/pc$^3$ (as for the Milky Way case, see table \ref{table:5.2}):
\begin{equation}
\sigma/m \gtrsim 10^{-18}  \rm{cm}^2/\rm{g} \, ,
\end{equation}
directly implying from our cross-section formula within the low energy approximation (\ref{eq:20}), that  $C_V \gtrsim 2 G_F$, that is the interaction strength can never be smaller than the weak interaction Fermi coupling.

\section{Discussion} \label{sec:out}

It is interesting to notice that the degenerate $\sim 10^1$~keV fermion core can reach core radii small enough to be suitable for the SgrA* observational constraints, as well as to reach BH-like compactness as in the case of larger elliptical galaxies. This alternative approach acquires special interest for ongoing and future observational campaigns (e.g. the BlackHole-Cam project\footnote{http://horizon-magazine.eu/space}), which would allow to verify the general relativistic effects expected in the surroundings of the central compact source in SgrA*; leading to a deeper scrutiny for the not-yet confirmed black hole hypothesis.
\begin{figure*}
\centering
\includegraphics[width=.49\hsize,clip]{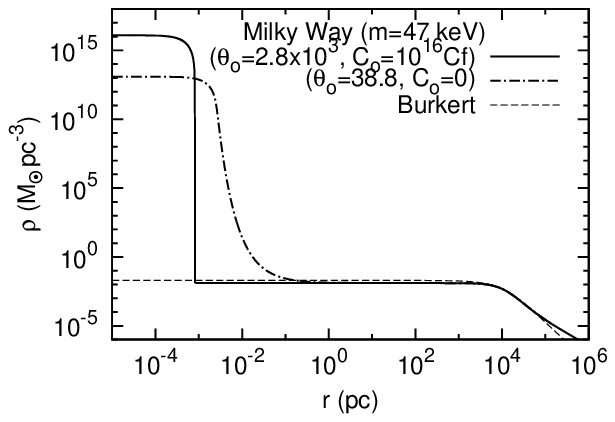}
\includegraphics[width=.49\hsize,clip]{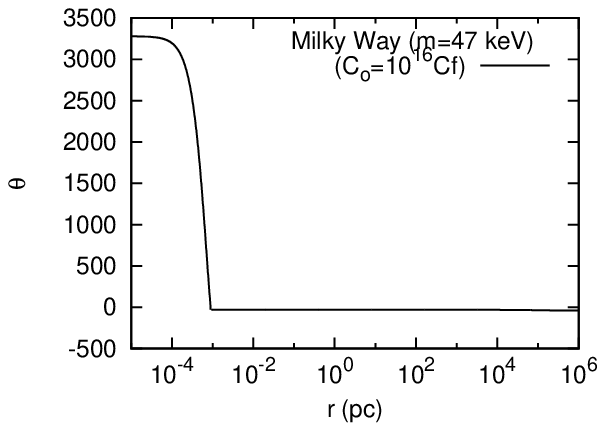}
\caption{\emph{Left}: mass density profiles for $m=47$~keV$/c^2$ in the interaction regime $\overline{C}_0=10^{16}$ where core and halo Milky Way observational constraints (\ref{eq:coreObs}--\ref{eq:haloObs}) are fulfilled, compared with the non-interacting case ($\overline{C}_0=0$) for the same ino mass in disagreement with the core observables. We also show for comparison the two parametric Burkert profile $\rho_B/[(1+r/h)(1+(r/h)^2)]$ with $\rho_B=2\times10^{-2} M_\odot/pc^3$ and $h=$10~kpc, which is the best DM halo fit of the Milky Way according to \citep{2009PASJ...61..153S}. \emph{Right}: degeneracy parameter profile in the interaction regime $\overline{C}_0=10^{16}$ for the same ino mass as in the Left panel.}\label{fig:2}
\end{figure*}

\begin{figure*}
\centering
\includegraphics[width=.49\hsize,clip]{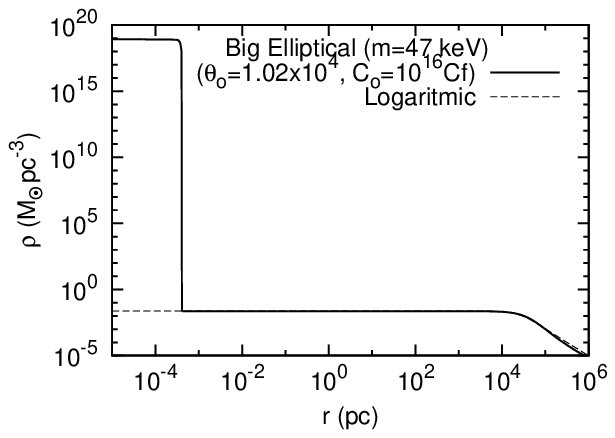}
\includegraphics[width=.49\hsize,clip]{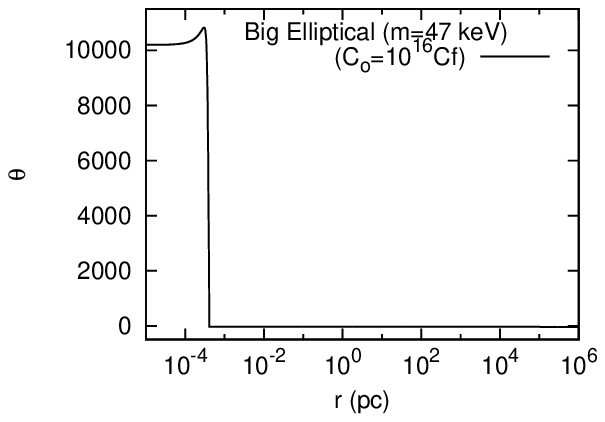}
\caption{\emph{Left}: mass density profiles for $m=47$~keV$/c^2$ in the same interaction regime as in figure~\ref{fig:2}, where the large elliptical core and halo observational constraints (\ref{eq:coreObsE}--\ref{eq:haloObsE}) are fulfilled. We also show for comparison the two-parameter Logaritmic profile $\rho_0^h[1+1/3(r/r_c^h)^2]/[1+(r/r_c^h)^2]^2$ with $\rho_0^h=2.0\times10^{-2} M_\odot/pc^3$ and $r_c^h=35$~kpc ($r_h\approx 2.4 r_c^h$), usually used in typical large elliptical galaxies~\citep{2011ApJ...729..129M}. \emph{Right}: degeneracy parameter profile in the interaction regime $\overline{C}_0=10^{16}$ for the same ino mass as in the left panel. Notice that the small bump in $\theta(r)$ around $r=r_c$ is absence in the non-interacting case, and thus it is originated from the presence of the meson-field in the system of equations (\ref{eq:eq1}--\ref{eq:tolman2})}.\label{fig:3}
\end{figure*}
In figures~\ref{fig:2} and \ref{fig:3} we show, for comparison, the overall density distribution $\rho(r)$ corresponding to the Milky Way and a large elliptical galaxy respectively, for a common self-interaction constant. For the Milky Way we show as well the $\rho(r)$ profile in the non-interacting case for the same ino mass $m=47$~keV$/c^2$. This comparison shows that, while in the non-interacting case ($\overline{C}_0=0$) the core observables (\ref{eq:coreObs}) \textit{are not} fulfilled, the presence of self-interactions allows to have higher degenerate cores satisfying both the core and halo Milky Way observables (\ref{eq:coreObs}--\ref{eq:haloObs}). It is important to notice that the density profile in the observationally well constrained halo region of figure~\ref{fig:2}, coincides with the one obtained in ref.~\cite{2015MNRAS.451..622R} in the absence of self-interactions for the same ino mass, and are in good agreement with the Burkert profile, best DM halo fit for the Milky Way, as shown in \cite{2009PASJ...61..153S}.

At this juncture we should point out that in our analysis above we did not discuss explicitly the r\^ole of baryonic matter, which of course is mainly concentrated through the central bulge and disk regions of galaxies. Its inclusion does not change the important conclusions of our work that the introduction of WDM fermion self interactions affects the core/halo structure and in particular induces higher central degeneracies and higher compactness of the inner quantum core. The key result presented here as well as in \cite{2015MNRAS.451..622R} is that the DM contribution is predominant in the inner core (below sub-pc scales), and in the halo region at the onset of the flat part of the given rotation curve; while in between baryonic matter prevails. Indeed, we can see from figure~\ref{fig:2} (left panel) that for the Milky Way, our model correctly predicts both the value and flattening of the circular velocity at distances $r\gtrsim 10$~kpc. A more complete quantitative analysis, including baryonic matter, is left for a future work.

We would also like to make one last, but not least, observation regarding the range of the self-interacting ino masses, $m \ge$ 47 ~keV/$c^2$. If we identify the inos with the (lightest) right-handed neutrino of the $\nu$MSM model~\cite{2005PhLB..631..151A,2009PThPh.122..185S,2009ARNPS..59..191B}, then the latter must have a very weak mixing angle with the SM lepton sector, and its mass must be less than 50~keV/$c^2$, otherwise the model would not be consistent with the current phenomenology, as can be seen from figure \ref{fig:DMsterile}. The above considerations, then, leave a very narrow range of the self-interacting `ino' mass $47 \le m \le 50$~keV/$c^2$, for the right-handed neutrino to play both a r\^ole as a WDM candidate and  a provider of a core-halo galactic structure in accordance to observation. Such constraints are of course alleviated if any mixing of the ino with the standard model sector is avoided, as done in the current article. Nevertheless, we find quite intriguing the fact that, starting from two entirely different approaches, one from particle physics, and the other from pure galactic astrophysics, one finds a consistent regime of `ino' masses within the WDM range. We believe that this is not a coincidence, and the aim of the current paper was to alert readers from these different communities to this important fact.

Finally, before closing, it is appropriate, as announced in the introduction of the article, to place our fermionic keV DM approach in context with the current state of affairs of cosmological DM and structure formation and some of the important open issues still faced by
the $\Lambda$CDM cosmology~\cite{2009NJPh...11j5029P}, such as: (i) the core-cusp problem~\cite{2010AdAst2010E...5D}, (ii) the ``lost satellite'' problem~\cite{2011PhRvD..83d3506P} and (iii) the, so-called, ``too big to fail'' problem~\cite{2011MNRAS.415L..40B,2012MNRAS.422.1203B}, the latter being a discrepancy between the most massive subhaloes arising within CDM and the dynamics of the brightest dSph galaxies of the Milky Way. All these problems have their root in the fact that cold DM particles have too short free streaming length during the epochs of galaxy formation, and therefore they form too clumped and too many structures than those observed. 

Our model provides a natural solution for (i) because the density profiles based on fermionic phase-space distributions develop always an extended plateau on halo scales (starting after the quantum core), in a way that resemble Burkert or cored Einasto profiles~\cite{2015MNRAS.451..622R} (see figures~\ref{fig:1} and \ref{fig:2}). Regarding the issues (ii) and (iii), it is important to bear in mind that our model does not directly deal with structure formation mechanisms, nor has employed (as yet) numerical simulations at such scales, and hence we are not in a position to make any concrete statement on these two issues. Nevertheless, for particle masses in the \emph{few} $10^{1-2}$ keV range  as obtained here, it has been extensively shown by now that the behaviour in the power spectrum (up to $\sim $ Mpc scales) is practically indistinguishable from that of standard CDM particles \cite{2009ARNPS..59..191B}, thus maintaining the expected results from large-scale structure observations.  There are issues, such as re-ionization, that we have still not examined, and thus at this stage we cannot make any concrete statements as to how much our model provides a substitute for CDM, although several of its features, as we have explained above, are indistinguishable from it. In fact, it may well be that there exist more than one DM species in the universe, and in this respect, our self interacting warm (right-handed neutr)``inos''  play an important r\^ole in the galactic-core structure, which was analyzed above, but a complete explanation/resolution of the large-scale structure problem in the cosmos may require synergies among different DM species, including CDM and WDM. More work is needed to arrive at firm conclusions in these matters.

 In this respect we mention for completeness that several proposals have been made recently towards a resolution of these issues within standard $\Lambda$CDM N-body simulations, including the self-interacting DM approach of \cite{2013MNRAS.430...81R}, as well as baryonic feedback processes~\cite{2012MNRAS.422.1231G,2013MNRAS.433.3539G}. Moreover, within the realm of (Newtonian) N-body simulations, $\Lambda$WDM cosmologies based on particles of \emph{few} keV have been developed to tackle the aforementioned $\Lambda$CDM discrepancies (i)-(iii)~\cite{2012MNRAS.420.2318L,2014MNRAS.439..300L}. Particles with such a small mass can suppress structure formation on small scales, due to their larger free streaming length caused by appreciable thermal velocities. Nevertheless, the fact that recent observations suggest that the number of Milky Way satellites is an order of magnitude greater than that predicted by WDM numerical simulations, casts doubts for the \emph{few} keV WDM scenario \cite{2008ApJ...688..277T}. Moreover, such light ($1-3$~keV) particles are also in strong tension with actual lower keV bounds set by current Ly-$\alpha$ forest constraints~\cite{2009PhRvL.102t1304B,2013PhRvD..88d3502V}.

All these issues present serious challenges for N-body simulation-based cosmologies (i.e $\Lambda$WDM in this case), associated with the `too warm' nature of the particles involved. Interestingly, the fact that the particle mass in our model is `colder' by a few keV compared to those WDM models, implies that our model does not suffer from such standard WDM problems. This, together with the fact that the model tackled successfully the important core-cusp discrepancy, as mentioned above, and that it avoids several of the undesired halo features characterizing the $\Lambda$CDM paradigm, offers significant support to the idea that our self-interacting model of right-handed neutrinos provides a physically important DM species, which may co-exist harmonically with other DM structures in the universe. In this respect, we believe that complementary searches for such keV right-handed neutrinos, either in neutrino oscillation experiments or elsewhere, are important.

\acknowledgments

The work of N.E.M. is supported in part by the London Centre for Terauniverse Studies (LCTS), using funding from the European Research Council via
the Advanced Investigator Grant 267352 and by STFC (UK) under the research grant ST/L000326/1. C.R.A and J.A.R are supported by the International Center for Relativistic Astrophysics Network (ICRANet). C.R.A also acknowledges the support from CONICET-Argentina. J.A.R acknowledges support from the International Cooperation Program CAPES-ICRANet financed by CAPES-Brazilian Federal Agency for Support and Evaluation of Graduate Education within the Ministry of Education of Brazil.

\bibliography{biblioDM}
\bibliographystyle{JHEP}

\appendix

\section{Central temperature parameters \label{app:A}}

A sufficiently precise determination of the central temperature parameter $\beta_0^h$ in the low relativistic regime of the model, when applied to normal galaxies, can be understood through the following two concepts (the value of the speed of light $c$ is here given in km/s):

1) Boltzmann regime at $r\sim r_m$: Since at $r\gtrsim r_m$ the degeneracy parameter fulfills $\theta(r)\ll-1$, the Fermi-Dirac statistics necessarily approaches the pure Boltzmann regime. The Boltzmann distribution function is characterized by the familiar one-dimensional velocity dispersion, $\sigma$, which is independent of the radius
\begin{equation}
\sigma^2=k_BT/m\, .
\label{eq:sigma}
\end{equation}

2) Classical isothermal-sphere condition. A classical self-gravitating system of Boltzmann-like particles in hydrostatic equilibrium is described by the isothermal-sphere model. The relation between the circular velocity $v_c(r)$ and $\sigma$ for an isothermal-sphere model is $v_c^2(r)=-\sigma^2(d\ln \rho(r)/d\ln r)$, where $\rho(r)$ is the mass density (see, e.g.,~\cite{2008gady.book.....B}). Different cored solutions to $\rho(r)$ depend only on the constant initial conditions $\rho_0^h$ and $\sigma$, implying a universal behavior (scaling) of the density profile. Thus, the logarithmic derivative evaluated at the halo radius $r_h$ (defined at the maximum of the velocity curve, \emph{i.e.} the onset of the flat part) is $(d\ln \rho(r)/d\ln r)|_{r_h}=-2.51$. This implies $v_h^2=2.51\sigma^2$, and, hence, using eq.~(\ref{eq:sigma}), one obtains
\begin{equation}
\beta_0^h=\frac{1}{2.51}\left(\frac{v_h}{c}\right)^2,
\end{equation}
which for $v_h=155$~km/s gives
\begin{equation}
\beta_0^h\equiv\beta(r_m)=1.065\times10^{-7}\, .
\end{equation}
This is the value we use in our phenomenological analysis of section \ref{sec:num}.

Finally, notice that, in order to obtain the central temperature parameters $\beta_0$ appearing in table \ref{table:5.2}, we  use the Tolman condition for isothermality $e^{\nu/2}T=constant$. The latter, together with the definition of $\beta=k_BT/(m c^2)$, implies  the relation (\ref{eq:tolman2}), which expresses
the temperature parameter at any given radius in terms of
the central temperature parameter $\beta_0$. For example, in case the massive quantum core of  SgrA* has a small compactness (i.e. $r_c\sim 6\times10^{-4}$~pc), as dictated by $GM_c/r_c\sim10^{-4}$, one necessarily has the following condition for the metric factor between core and halo $e^{\frac{\nu_0-\nu(r_m)}{2}}\approx1$. Consequently, the following relation can be established $\beta_0^h=\beta_0$ at three-digit precision. Instead, in cases with higher core compactness (i.e. when $r_c\sim 10^{-6}$~pc in the Milky Way case, or when $r_c\sim 10^{-5}$~pc in the elliptical galaxy case) as shown in table \ref{table:5.2}, slightly higher values of $\beta_0$ are obtained by the use of the Tolman condition.

\section{Total cross-section $\sigma^{tot}$ within a four-Majorana-fermion (elastic scattering) interaction}\label{app:B}

A scattering process consisting in two incident particles with four-momentum ($p_i=(E_i,\bf{p_i})$), $i=1,2$, which collide and produce two final particles with momentum ($p'_i=(E'_i,\bf{p'_i})$), $i=1,2$ (all with definite polarization/spin states) has a total cross-section in the center-of-mass (CoM) system given by (see eqn. (8.19) in \cite{mandl1993quantum})

\begin{equation}
\sigma^{tot}_{CoM}=\int_0^1 d(cos \theta) \int_0^{2\pi} d\phi\left(\frac{d\sigma}{d\Omega}\right)_{CoM}=\frac{1}{2}\int d\Omega\left(\frac{d\sigma}{d\Omega}\right)_{CoM} \, .
\label{eq:1}
\end{equation}

The integration in the last member is made over the complete solid angle, while the integrals involving the ($\theta,\phi$) scattering angles were performed over the forward hemisphere ($0\leq\theta\leq\pi/2$) corresponding to physically distinguishable events, and proper for a process with two identical particles in the final state, as it is our case of interest. With $(d\sigma/d\Omega)_{CoM}$ the differential cross-section in the Center of Mass system given by the general formula \cite{mandl1993quantum}.
\begin{equation}
\left(\frac{d\sigma}{d\Omega}\right)_{CoM}=\frac{1}{64\pi^2(E_1+E_2)^2}\frac{|\bf{p'_i}|}{|\bf{p_i}|}\,\Pi(2 m_l)\,|\mathcal{M}|^2 \, ,
\label{eq:2}
\end{equation}
where the sub-index l runs over all external leptons in the scattering process, and $\mathcal{M}$ is the Feynman amplitude of the process.

\begin{figure}
\centering
\includegraphics[width=0.5\textwidth]{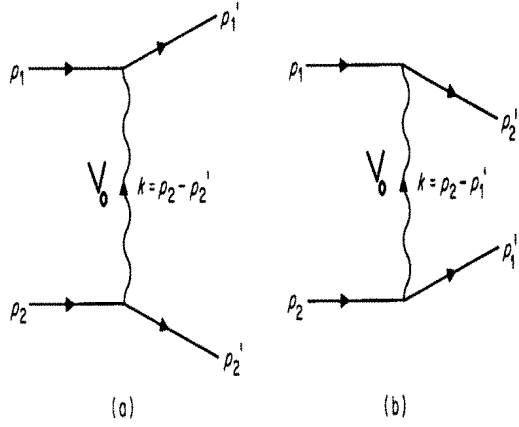}
\caption{Feynman diagrams for $\nu_s$ - $\nu_s$ scattering, with $V_0$ the massive-vector boson mediator}\label{fig:vrotfamily}
\end{figure}

A lepton-lepton ($\nu_s$ - $\nu_s$ in this case) scattering process has two possible Feynman diagrams (differing by the interchange of the final lepton states {$1'\leftrightarrow 2'$}) given in figure 1, with two different associated Feynman amplitudes $\mathcal{M}_a$ and $\mathcal{M}_b$, built up below by the use of the Feynman rules associated to each diagram (a) and (b) respectively (see e.g.~\cite{mandl1993quantum}, appendix~\ref{app:B}). It is important to realize that being the sterile neutrinos neutral (no electric charge), the only possibility for the boson mediator in the interaction (at lowest order in the interaction term of the Lagrangian), is to be a neutral vector boson $V_0$. This reasoning is implemented in complete analogy to the scattering of leptons within electroweak theory, where the $Z_0$ boson can be viewed as the analogue of our dark-sector $V_0$ boson.

The Feynman amplitudes $\mathcal{M}_a$ and $\mathcal{M}_b$ read, in accordance with the corresponding Feynman rules, as follows

\begin{eqnarray}
&\mathcal{M}_a=-\frac{g_V^2}{16 \cos^2(\theta'_W)}[\bar{u}_{s_1}(\mathbf{p}'_1) \gamma^\alpha(1-\gamma_5)u_{r_1}(\mathbf{p}_1)] iD_{F_{\alpha\beta}}(k=p_2-p'_2,m)[\bar{u}_{s_2}(\mathbf{p}'_2) \gamma^\alpha(1-\gamma_5)u_{r_2}(\mathbf{p}_2) ] \nonumber\\
&\mathcal{M}_b=\frac{g_V^2}{16 \cos^2(\theta'_W)}[\bar{u}_{s_2}(\mathbf{p}'_2) \gamma^\alpha(1-\gamma_5)u_{r_1}(\mathbf{p}_1)] iD_{F_{\alpha\beta}}(k=p_2-p'_1,m)[\bar{u}_{s_1}(\mathbf{p}'_1) \gamma^\alpha(1-\gamma_5)u_{r_2}(\mathbf{p}_2) ] \nonumber
\label{eq:3}
\end{eqnarray}

Where $g_V$ is the Yukawa coupling constant of our (only) interaction term in the lagrangian ($\mathcal{L}_I$), with $\theta'_W$ the `pseudo-weak' mixing angle\footnote{This can be formally implemented within our (minimally extended) beyond SM sterile neutrino, through the linear combinations (dark sector fields) of $B_\mu(x)$ (e.g. \cite{mandl1993quantum} Eqtn. (12.45)), in total analogy with the electroweak mixing angle $\theta_W$.}, $u_r(\textbf{p})$ ($u_s(\textbf{p'})$) the four-Majorana spinor corresponding to a particle of momentum $\textbf{p}$, positive energy E, and spin indexes $r=1,2$ ($s=1,2$) of the neutrinos in the initial (and final) states respectively (being $\bar{u}$ the adjoint). And $D_{F_{\alpha\beta}}(k,m)$ the massive-vector boson propagator which reads

\begin{equation}
D_{F_{\alpha\beta}}(k,m)=\frac{-g_{\alpha\beta}+k_\alpha k_\beta/m_V^2}{k^2-m_V^2+i\epsilon}\, ,
\label{eq:4}
\end{equation}
with $g_{\alpha\beta}$ the metric tensor with signature mostly minus and locally Minkowskian (equivalence principle), $m_V$ the vector boson mass and $\epsilon$ a very small parameter (to avoid pole divergences). As it will be clear later, the typical sterile neutrino (one particle) energies we are dealing in our galactic environment, implies we are in a `low energy limit', and we can safely assume $\bf{k^2\ll m_V^2}$ (for `normally' heavy bosons $V_0$ with masses above keV, as we assume here). With this inequality we have for (\ref{eq:4})

\begin{eqnarray}
D_{F_{\alpha\beta}}(k,m) \rightarrow \frac{g_{\alpha\beta}}{m_V^2} \qquad (k^2/m_V^2\rightarrow 0)\, ,
\label{eq:5}
\end{eqnarray}
which leads to the following simplifications in the Feynman amplitudes $\mathcal{M}_a$ and $\mathcal{M}_b$
\begin{eqnarray}
&\mathcal{M}_a=-i\left(\frac{g_V}{m_V}\right)^2\frac{1}{16 \cos^2(\theta'_W)}[\bar{u}_{s_1}(\mathbf{p}'_1) \gamma^\alpha(1-\gamma_5)u_{r_1}(\mathbf{p}_1)] [\bar{u}_{s_2}(\mathbf{p}'_2) \gamma_\alpha(1-\gamma_5)u_{r_2}(\mathbf{p}_2) ] \nonumber\\
&\mathcal{M}_b=i\left(\frac{g_V}{m_V}\right)^2\frac{1}{16 \cos^2(\theta'_W)}[\bar{u}_{s_2}(\mathbf{p}'_2) \gamma^\alpha(1-\gamma_5)u_{r_1}(\mathbf{p}_1)] [\bar{u}_{s_1}(\mathbf{p}'_1) \gamma_\alpha(1-\gamma_5)u_{r_2}(\mathbf{p}_2) ] \nonumber
\label{eq:6}
\end{eqnarray}

The Feynman amplitude factor in (\ref{eq:2}) can be written now in terms of the above amplitudes $|\mathcal{M}|^2=|\mathcal{M}_a+\mathcal{M}_b|^2=|\mathcal{M}_a|^2+|\mathcal{M}_b|^2+\mathcal{M}_a\mathcal{M}^*_b+(\mathcal{M}_a\mathcal{M}_b)^*$. With $^*$ the complex conjugate.

In following, we assume that the `polarization' state of our (initial and final) neutrinos is not known, and therefore we shall calculate the (standard) `unpolarized' cross-section, for which we must \textit{average} $|\mathcal{M}|^2$ over all initial and final lepton spins. Indeed, notice that our Majorana singlets $N_{R1}$ are of right-handed chirality, but nevertheless they may have both positive and negative helicities. This averaging procedure consists in average over initial spin states (i.e. $1/4\sum_{r1}\sum_{r2}$), and summing over final spin states ($\sum_{s1}\sum_{s2}$) (see e.g. \cite{mandl1993quantum}, section 8.2).
Therefore, each Feynman amplitude term ($|\mathcal{M}_a|^2$,$|\mathcal{M}_b|^2$,etc.) is now replaced by

\begin{equation}
X=\frac{1}{4}\sum_{r1}\sum_{r2}\sum_{s1}\sum_{s2}|\mathcal{M}(r1,r2,s1,s2)|^2 \, ,
\end{equation}

leading to a differential cross-section (\ref{eq:2}) of the form

\begin{eqnarray}
\left(\frac{d\sigma}{d\Omega}\right)_{CoM}\propto(X_{aa}+X_{bb}+X_{ab}+X^*_{ab}) \, ,
\label{eq:7}
\end{eqnarray}

where

\begin{eqnarray}
X_{aa}&=\frac{1}{4}\sum_{spins}|\mathcal{M}_a|^2 \\
X_{bb}&=\frac{1}{4}\sum_{spins}|\mathcal{M}_b|^2 \\
X_{ab}&=\frac{1}{4}\sum_{spins}\mathcal{M}_a\mathcal{M}_b^*
\end{eqnarray}

By inserting equations (\ref{eq:6}) in the above Feynman amplitude factors, we have correspondingly

\begin{align*}
X_{aa}=C\sum_{r1}\sum_{s1}{[\bar{u}_{s_1}(\mathbf{p}'_1) \gamma^\alpha(1-\gamma_5)u_{r_1}(\mathbf{p}_1)]^2}\sum_{r2}\sum_{s2}{[\bar{u}_{s_2}(\mathbf{p}'_2) \gamma_\alpha(1-\gamma_5)u_{r_2}(\mathbf{p}_2)]^2} \\
X_{bb}=C\sum_{r1}\sum_{s2}{[\bar{u}_{s_2}(\mathbf{p}'_2) \gamma^\alpha(1-\gamma_5)u_{r_1}(\mathbf{p}_1)]^2}\sum_{r2}\sum_{s1}{[\bar{u}_{s_1}(\mathbf{p}'_1) \gamma_\alpha(1-\gamma_5)u_{r_2}(\mathbf{p}_2)]^2} \\
X_{ab}=-C\sum_{r1}\sum_{r2}\sum_{s1}\sum_{s2}[\bar{u}_{s_1}(\mathbf{p}'_1) \gamma^\alpha(1-\gamma_5)u_{r_1}(\mathbf{p}_1)][\bar{u}_{s_2}(\mathbf{p}'_2) \gamma_\alpha(1-\gamma_5)u_{r_2}(\mathbf{p}_2)]x\\ [\bar{u}_{s_2}(\mathbf{p}'_2) \gamma^\alpha(1-\gamma_5)u_{r_1}(\mathbf{p}_1)][\bar{u}_{s_1}(\mathbf{p}'_1) \gamma_\alpha(1-\gamma_5)u_{r_2}(\mathbf{p}_2)]\, ,
\end{align*}

with $C=(g_V/m_V)^4\,1/[4^5\cos^4(\theta'_W)]$. Then, by using the Hermiticity condition of the gamma matrices ($(\gamma^\mu)^\dag=\gamma^0\gamma^\mu\gamma^0$) and the regrouping spinor-component technics (e.g. \cite{mandl1993quantum}, section 8.2), it can be shown that the Feynman amplitude factors written above can be easily expressed in terms of traces of products of gamma matrices as follows:

\begin{align*}
&X_{aa}=C\, \rm{\textbf{Tr}}[\Lambda_M^+(\mathbf{p'_1})\gamma^\alpha(1-\gamma^5)\Lambda_M^+(\mathbf{p_1})\gamma^\beta(1-\gamma^5)]\, \rm{\textbf{Tr}}[\Lambda_M^+(\mathbf{p'_2})\gamma_\alpha(1-\gamma^5)\Lambda_M^+(\mathbf{p_2})\gamma_\beta(1-\gamma^5)]\\
&X_{bb}=C\, \rm{\textbf{Tr}}[\Lambda_M^+(\mathbf{p'_2})\gamma^\alpha(1-\gamma^5)\Lambda_M^+(\mathbf{p_1})\gamma^\beta(1-\gamma^5)]\, \rm{\textbf{Tr}}[\Lambda_M^+(\mathbf{p'_1})\gamma_\alpha(1-\gamma^5)\Lambda_M^+(\mathbf{p_2})\gamma_\beta(1-\gamma^5)]\\
&X_{ab}=-C\, \rm{\textbf{Tr}}[\Lambda_M^+(\mathbf{p'_1})\gamma^\alpha(1-\gamma^5)\Lambda_M^+(\mathbf{p_1})\gamma^\beta(1-\gamma^5)
\Lambda_M^+(\mathbf{p'_2})\gamma_\alpha(1-\gamma^5)\Lambda_M^+(\mathbf{p_2})\gamma_\beta(1-\gamma^5)]
\end{align*}

where $\Lambda_M^+(\textbf{p})\equiv\Lambda_{\alpha\beta}^+(\textbf{p})=\sum_{r=1}^2 u_{r\alpha}(\textbf{p})\bar{u}_{r\beta}(\textbf{p})=\frac{-\slashed{p}+m/2}{m}$ is the Majorana (positive) energy projection operator (see next section for details and difference w.r.t the Dirac projector), with $\slashed{p}=\gamma^\lambda p_\lambda$ .In next we rewrite the above traces as $X_{aa}=(C/m^4)A^{\alpha\beta}\,B_{\alpha\beta}$ and $X_{bb}=(C/m^4)E^{\alpha\beta}\,F_{\alpha\beta}$ and proceed to calculate each factor-element $A,B,E,F$.
\begin{equation}
A^{\alpha\beta}=\textbf{Tr}[(-\slashed{p}'_1+m/2)\gamma^\alpha(1-\gamma^5)(-\slashed{p}_1+m/2)\gamma^\beta(1-\gamma^5)]
\label{eq:8}
\end{equation}
By using the following properties of the $\gamma$-matrices: \textbf{i)} Tr$[\gamma^\alpha \gamma^\beta \gamma^\gamma ..]=0$ (with $[\gamma^\alpha \gamma^\beta \gamma^\gamma ..]$) an odd number of $\gamma$ matrices; \textbf{ii)} Tr$[\gamma^5\gamma^\alpha\gamma^\beta]=$Tr$[\gamma^5\gamma^\alpha\gamma^\beta\gamma^\gamma]=0$; \textbf{iii)} $[\gamma^\mu,\gamma^5]_+=0$ (with $[\,]_+$ the anitconmutator); \textbf{iv)} $(\gamma^5)^2=1$. Then, equation (\ref{eq:8}) reads
\begin{equation}
A^{\alpha\beta}=2 p'_{1_{\lambda}}p_{1_{\gamma}}\textbf{Tr}[\gamma^\lambda\gamma^\alpha\gamma^\gamma\gamma^\beta-\gamma^5\gamma^\lambda\gamma^\alpha\gamma^\gamma\gamma^\beta]\, ,
\label{eq:9}
\end{equation}
where $\slashed{p}'_1=\gamma^\lambda p'_{1_{\lambda}}$ and $\slashed{p}_1=\gamma^\gamma p_{1_{\gamma}}$. So, by using the following properties \textbf{v)} Tr$[\gamma^\lambda\gamma^\alpha\gamma^\gamma\gamma^\beta]=4(g^{\lambda\alpha}g^{\gamma\beta}-g^{\lambda\gamma}g^{\alpha\beta}+g^{\lambda\beta}g^{\alpha\gamma})$; \textbf{vi)} Tr$[\gamma^5\gamma^\lambda\gamma^\alpha\gamma^\gamma\gamma^\beta]=-4i\epsilon^{\lambda\alpha\gamma\beta}$, we finally have
\begin{equation}
A^{\alpha\beta}=8 (p'^\alpha_1p^\beta_1-p'^\gamma_1p_{1_\gamma}g^{\alpha\beta}+p^\alpha_1p'^\beta_1)+8i p'_{1_\lambda}p_{1_\gamma}\epsilon^{\lambda\alpha\gamma\beta} \,.
\label{eq:10}
\end{equation}
Analogously as done for the $A$ factor, the $B$ factor reads (using \textbf{i-vi}) and recalling $g_{\alpha\delta}\gamma^{\delta}=\gamma_\alpha$, etc)
\begin{equation}
B_{\alpha\beta}=8 (p'_{2_\alpha}p_{2_\beta}-p'^\mu_2p_{2_\mu}g_{\alpha\beta}+p_{2_\alpha}p'_{2_\beta})+8i p'^\lambda_2p^\gamma_2\epsilon_{\lambda\alpha\gamma\beta} \, ,
\label{eq:11}
\end{equation}
where $\slashed{p}'_2=\gamma^\mu p'_{2_{\mu}}$ and $\slashed{p}_2=\gamma^\nu p_{2_{\nu}}$. Finally, by multiplying Eqtns. (\ref{eq:10}) and (\ref{eq:11}) we have
\begin{equation}
A^{\alpha\beta}B_{\alpha\beta}=64[26(p'_1p'_2)(p_1p_2)-3(p'_1p_1)(p_2p'_2)+2(p'_1p_2)(p_1p'_2)] \, ,
\label{eq:12}
\end{equation}
where $p'^\mu_i p_{i_\mu}\equiv p_i p'_i$ ($i=1,2$), and we have used the important properties of the Levi-Civita symbols (or alternating symbols), \textbf{vii)} $\epsilon^{\lambda\alpha\gamma\beta}\epsilon_{\lambda\alpha\gamma\beta}=-24$; and \textbf{viii)} $\epsilon_{\lambda\mu\nu\pi}=\pm1$, with the $+$ ($-$) sign is for $(\lambda,\mu,\nu,\pi)$ and even (odd) permutation of $(0,1,2,3)$, and vanishes if two or more indices are the same. This last property was used to show that the crossing-products in $A$ (\ref{eq:10}) times $B$ (\ref{eq:11}), are $0$, and only two terms survive leading to (\ref{eq:11}).

We proceed with the calculation of the factor-elements $E$ and $F$, and notice that $E^{\alpha\beta}=_{\slashed{p}'_1\rightarrow\slashed{p}'_2} A^{\alpha\beta}$, and $F_{\alpha\beta}=_{\slashed{p}'_2\rightarrow\slashed{p}'_1} B_{\alpha\beta}$. Therefore, it leads straightforwardly to
\begin{eqnarray}
E^{\alpha\beta}=8 (p'^\alpha_2p^\beta_1-p'^\gamma_2p_{1_\gamma}g^{\alpha\beta}+p^\alpha_1p'^\beta_2)+8i p'_{2_\lambda}p_{1_\gamma}\epsilon^{\lambda\alpha\gamma\beta} \\ \nonumber
F_{\alpha\beta}=8 (p'_{1_\alpha}p_{2_\beta}-p'^\mu_1p_{2_\mu}g_{\alpha\beta}+p_{2_\alpha}p'_{1_\beta})+8i p'^\lambda_1p^\gamma_2\epsilon_{\lambda\alpha\gamma\beta} \, ,
\label{eq:13}
\end{eqnarray}
leading directly to (after using \textbf{vii-viii})
\begin{equation}
E^{\alpha\beta}F_{\alpha\beta}=64[26(p'_1p'_2)(p_1p_2)-3(p'_1p_2)(p_1p'_2)+2(p'_1p_1)(p_2p'_2)] \, .
\label{eq:14}
\end{equation}

With the results obtained up to here from equations (\ref{eq:12}) and (\ref{eq:14}), plus the use of the $\nu_s$ - $\nu_s$ elastic scattering kinematics detailed in the last section below, we finally have for $X_{aa}$ and $X_{bb}$
\begin{eqnarray}
X_{aa}=\left(\frac{g_V}{m_V}\right)^4\frac{1}{4^2m^4\cos^4(\theta'_W)}[26(p_1p_2)^2-3(p_1p'_1)^2+2(p_1p'_2)^2] \\
X_{bb}=\left(\frac{g_V}{m_V}\right)^4\frac{1}{4^2m^4\cos^4(\theta'_W)}[26(p_1p_2)^2-3(p_1p'_2)^2+2(p_1p'_1)^2] \, .
\label{eq:15}
\end{eqnarray}

We now calculate the last term $X_{ab}$, which can be written as traces of products of up to nine $\gamma$ matrices (after having used all the properties \textbf{i-vi})
\begin{equation}
X_{ab}=(C/m^4)\,8\rm{\textbf{Tr}}[\slashed{p}'_1\gamma^\alpha\slashed{p}_1\gamma^\beta\slashed{p}'_2\gamma_\alpha\slashed{p}'_2\gamma_\beta- \gamma^5\slashed{p}'_1\gamma^\alpha\slashed{p}_1\gamma^\beta\slashed{p}'_2\gamma_\alpha\slashed{p}'_2\gamma_\beta]\, .
\label{eq:16}
\end{equation}
By the use of the following properties in equation (\ref{eq:16}): \textbf{ix)} Tr$[UV]=$Tr$[VU]$ (with $U,V$ $4$x$4$ matrices), and the following contraction properties of $\gamma$-matrices \textbf{x)} $\gamma_\lambda\gamma^\alpha\gamma^\beta\gamma^\gamma\gamma^\lambda=-2\gamma^\gamma\gamma^\beta\gamma^\alpha$; \textbf{xi)} $\gamma_\lambda\slashed{P}\slashed{Q}\gamma^\lambda=4PQ$ (with $P$ and $Q$ four-vectors); it can be easily shown the following result (coming from the first trace term in (\ref{eq:16}) only, while the second trace term containing $\gamma^5$ vanishes by the use of the extra-property \textbf{ii})
\begin{equation}
X_{ab}=-(C/m^4)4^3(p_2p_1)\rm{Tr}[\slashed{p}'_1\slashed{p}'_2]=-\left(\frac{g_V}{m_V}\right)^4\frac{1}{4m^4\cos^4(\theta'_W)}(p_1p_2)^2\, ,
\label{eq:17}
\end{equation}
where for the right hand side in the above equation we have used the property \textbf{xii)} Tr$[\gamma^\alpha\gamma^\beta]=4g^{\alpha\beta}$, and the kinematical properties given in the last section.

We note that $X_{ab}$ is real, and therefore we have $X^*_{ab}=X_{ab}$. This last result together with (\ref{eq:15}) and (\ref{eq:17}) implies finally for the differential cross-section (\ref{eq:7}), the following equation (we have used the elastic scattering condition $\textbf{p}'_1=\textbf{p}_1$, and the assumption of typical initial sterile neutrino energies ($E_1\approx E_2\equiv E$)
\begin{equation}
\left(\frac{d\sigma}{d\Omega}\right)_{CoM}=\left(\frac{g_V}{m_V}\right)^4\frac{1}{4^4\pi^2E^2\cos^4(\theta'_W)}[60(p_1p_2)^2-(p_1p'_1)^2-(p_1p'_2)^2]\, ,
\label{eq:18}
\end{equation}
which, by the use of the kinematics for this process in terms of the typical particle energy $E^2=(p^2+m^2)$, the three-momentum $p$, and the scattering angle $\theta$ (see last section for details), we have finally for the differential cross-section
\begin{equation}
\left(\frac{d\sigma}{d\Omega}\right)_{CoM}=\left(\frac{g_V}{m_V}\right)^4\frac{2}{4^4\pi^2\cos^4(\theta'_W)}[29E^2+60p^2+p^4/E^2(30-\cos^2(\theta))]\, ,
\label{eq:19a}
\end{equation}
leading directly by the use of equation (\ref{eq:1}) the total cross-section for our $\nu_s$ - $\nu_s$ scattering process in terms of $p$ and $m$
\begin{equation}
\sigma^{tot}_{CoM}=\left(\frac{g_V}{m_V}\right)^4\frac{1}{4^3\pi^2\cos^4(\theta'_W)}[29m^2+89p^2+89/3\frac{p^4}{p^2+m^2}]\, .
\label{eq:19}
\end{equation}
This is the expression we use in section 3.3 [eqn.~(\ref{eq:totalsigma})]. In fact, for our purposes in this work it suffices to consider only a single Abelian vector interaction, thus ignoring any mixing angle in the dark sector. This implies that we can safely set $\theta_W' = 0$.

\subsection*{Majorana equation, plane wave solutions and energy projection operator}

As we have shown in section \ref{sec:rhn} of the paper, the equation of motion for our right-handed sterile neutrinos singlets $N_{R1}$ is given by
\begin{equation}
\bar{N}_{R1}(\slashed{p}+1/2 m)=0\, ,
\label{eq:21}
\end{equation}
with $\slashed{p}=i\gamma^\mu D_\mu$, being (\ref{eq:21}) obtained after taking the variations of the corresponding Majorana Lagrangian density $\mathcal{L}_{N_{R1}}$ w.r.t $N_{R1}$, given by equation (2.3) of the paper. Equation (\ref{eq:21}) admits plane wave solutions of the form
\begin{equation}
N_{R1}(x)=const.(u_r(\mathbf{p}))e^{-ipx}
\end{equation}
where $u_r(\textbf{p})$ are the constant singlet four-spinors with positive energy $E$ and momentum $\textbf{p}$.

Now, if we define a Majorana (positive) energy projector operator of the form
\begin{equation}
\Lambda^+(\mathbf{p})=\frac{\slashed{p}-1/2m}{-m}=\frac{-\slashed{p}+1/2m}{m}\, ,
\label{eq:22}
\end{equation}
it can be directly seen that fulfills with the property of projecting out the positive energy solutions from a linear combination of possible plane wave states $u_r(\textbf{p})$ and $v_r(\textbf{\textbf{p}})$, with $v_r$ a negative energy plane wave state (if it would exist), i.e.
\begin{equation}
\bar{u}_r(\mathbf{p})\Lambda^+(\mathbf{p})=\bar{u}_r(\mathbf{p}); \qquad \bar{v}_r(\mathbf{p})\Lambda^+(\mathbf{p})=0\, ,
\label{eq:23}
\end{equation}
where $\bar{u}_r(\textbf{p})=u^\dag_r(\textbf{p})\gamma^0$ is the adjoint, and $u^\dag_r$ the transpose spinor. Therefore, with (\ref{eq:23}), and by the use of the completeness relation fulfilled by the constant-spinors: $\Sigma_r[u_{r\alpha}(\textbf{p})\bar{u}_{r\beta}(\textbf{p})-v_{r\alpha}(\textbf{p})\bar{v}_{r\beta}(\textbf{p})]=\delta_{\alpha\beta}$ (see also \cite{mandl1993quantum}, appendix A.4), it follows the following relation for our projector operator\footnote{Notice that this was done, in complete analogy with the Dirac case, where the projector was defined by $\Lambda^+_D=(\slashed{p}+m)/2m$, in accordance with the Dirac equation $\bar{u}_r(\textbf{p})(\slashed{p}-m)=0$, and following the same properties shown here, among others (\cite{mandl1993quantum}, appendix A.4)}.
\begin{equation}
\Lambda^+(\mathbf{p})\equiv\Lambda^+_{\alpha\beta}(\mathbf{p})=\sum_{r=1}^2 u_{r\alpha}(\mathbf{p})\bar{u}_{r\beta}(p)\, ,
\end{equation}
which will be largely used in our cross-section calculations, allowing us to calculate the $X_{aa},X_{bb},X_{ab}$ terms.

\subsection*{$\nu_s$ - $\nu_s$ elastic scattering kinematics in the CoM system}

The kinematics of an elastic scattering process in the Center of Mass system is given by the following equations, accordingly with figure 2
\begin{figure}
\centering
\includegraphics[width=0.5\textwidth]{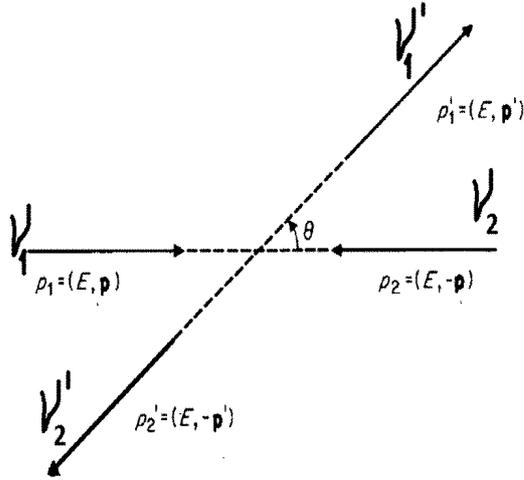}
\caption{Kinematics for a $\nu_s$ - $\nu_s$ scattering process in the CoM system.}
\end{figure}

\begin{eqnarray}
p_1=(E,\mathbf{p}_1);  \qquad  p'_1=(E,\mathbf{p}'_1) \\ \nonumber
p_2=(E,-\mathbf{p}_1);  \qquad  p'_2=(E,-\mathbf{p}'_1)
\end{eqnarray}
where the equalities $\textbf{p}_1=-\textbf{p}_2$ and $\textbf{p}'_1=-\textbf{p}'_2$ implies we are in the Center of Mass (CoM) system. Moreover, as we are under the assumption of elastic scattering conditions, the following equality for the three-momentum holds: $p=p'$, with $p\equiv|\textbf{p}|$ and $p'\equiv|\textbf{p}'|$. Finally, we have used the assumption that the initial particle Energies are of the same order, i.e. $E_1=E_2\equiv E$. Then we must have
\begin{eqnarray}
p_1p'_1=p_2p'_2=E^2-p^2\cos(\theta) \\ \nonumber
p_1p'_2=p_2p'_1=E^2+p^2\cos(\theta) \\ \nonumber
p_1p_2=p'_1p'_2=E^2+p^2 \nonumber
\end{eqnarray}

\section{Effective four-right-handed-Majorana-neutrino interaction coupling in the $\nu$MSM}\label{app:C}

\begin{figure}
\centering
\includegraphics[width=0.5\textwidth]{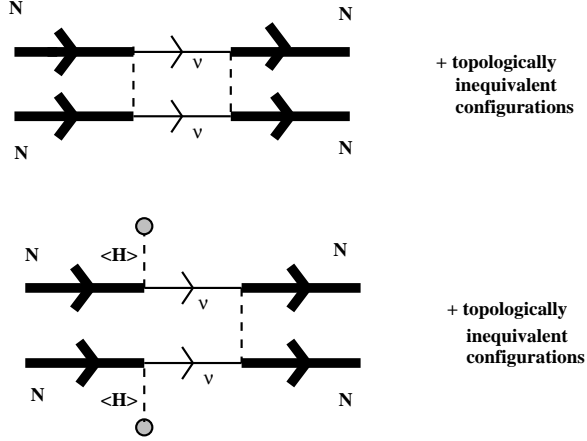}
\caption{Generic Feynman diagrams for the effective four-fermion right-handed Majorana neutrino coupling, in the context of $\nu$MSM.
The topologically inequivalent graph configurations are not shown explicitly as we are only interested in an order of magnitude estimate. Thick (thin) continuous lines indicate right-handed (active) Neutrinos N ($\nu$). Dashed lines indicate the Higgs (H) field, and the grey blobs denote the Higgs vacuum expectation value $\langle H \rangle \ne 0$. Upper pannel: One-loop, cut-off dependent diagram. Lower pannel: Tree-level diagram due to Majornana right-handed external line transmulation into an active (light) neutrino and Higgs field disappearance into the vacuum.}\label{fig:nmsmgraphs1}
\end{figure}

In this appendix we would like to explain why the induced four-fermion Majorana right-handed neutrino terms are subleading compared to the strong self-interactions considered above. This stems from the very weak nature of the Yukawa couplings $F_{\alpha I}$ as dictated by the seesaw mechanism which is assumed to be in operation here~\cite{2009PhRvL.102t1304B} so as to give a mass in the active neutrinos,
via their mixing with the sterile (Majorana) neutrinos of Mass $M_I =1, 2,3 $,  of the form:
\begin{equation} m_{\alpha \, \nu} = F^2_{\alpha I} \, \frac{v^2}{M_I}~, \, \quad \alpha = e, \mu, \tau
\label{seesaw}
\end{equation}
where $v \sim 175 $ GeV is the Higgs v.e.v. and the mass of the heavy neutrinos  is assumed much larger than the induced active neutrino mass $m_{\alpha \nu}$. In the $\nu$MSM one has the mass hierarchy~\cite{2005PhLB..631..151A}
$$ M_{1} = O(10-50 ~{\rm keV}) \ll M_{2}=M_{3} = O(1 ~{\rm GeV}) $$
and thus, assuming as a representative range (upper bound) of active neutrino masses the order of magnitude implied by the atmospheric experiments (which is in agreement also with the cosmological bound),
$$ m_\nu^2 \sim  |\Delta m_{\rm atm} |^2
= (2.40 + 0.12 - 0.11) \times 10^{-3}~{\rm eV}^2~,$$
we obtain from (\ref{seesaw})~~\cite{2009ARNPS..59..191B}:
\begin{equation}\label{f1v}
F_{\alpha 1} \sim 10^{-10}, \,  \quad F_{\alpha 2,3} \sim 10^{-7},
\end{equation}
thereby pointing towards very weak Yukawa couplings as stated previously. It should be noted though that it suffices to generate masses for three active neutrinos via (\ref{seesaw}) by assuming that the only non-vanishing Yukawa couplings are $F_{\alpha 2,3}$, in which case the mixing of the lightest neutrino $N_1$ with the matter sector is suppressed.

We next notice that the couplings of the Majorana neutrinos to the observable sector would itself yield an effective four Majorana neutrino
contact interactions, via two kind of processes:
\begin{itemize}

\item{(i)} through an Ultraviolet-momentum-cut-off-$\Lambda$ dependent  one particle irreducible box (one-loop) diagram (to leading contribution in perturbation expansion) with an active neutrino (assumed massless for all practical purposes) and a Higgs as internal lines, indicated in the upper panel of figure~\ref{fig:nmsmgraphs1}. Concentrating on the lightest right-handed neutrino for concreteness (thus assuming $F_{\alpha 1} \ne 0$ (cf. (\ref{f1v})), the corresponding four - $N_1$-neutrino scattering amplitude that would generate the contact interaction in an effective field theory framework would be proportional to
\begin{equation}\label{ff}
(F_{\alpha 1})^4 \times \int \frac{d^4 k}{(2\pi)^4} \frac{1}{\slashed{k}} \, \frac{1}{\slashed{p_3} - \slashed{p_1} + \slashed{k}} \, \frac{1}{(p_2 + k)^2 - M_H^2}
\, \frac{1}{(p_1 - k)^2 - M_H^2}~,
\end{equation}
where polarization spinors have been omitted for brevity, $p_{1,2}$ denote incident Majorana neutrino external momenta, whilst $p_{3,4}$ the outgoing ones, and $M_H$ is the mass of the Higgs field (assumed massive in the galactic era we are interested in, since the electroweak symmetry is broken at that epoch.

\item{(ii)} processes indicated generically in the lower panel of figure.~\ref{fig:nmsmgraphs1}, which involve the transmutation of an incident (or outgoing) Right-handed Majorana neutrino line into an active neutrino one, via a disapperance of a Higgs line into the vacuum. Such graphs contribute an effective four-Majorana Neutrino, whose amplitude is of order (up to numerical factors counting the various inequivalent graphs and ignoring spinor polarizations for brevity)
\begin{equation}\label{tlec}
(F_{\alpha 1})^4 \frac{v^2}{(p_1-p_3)^2 - M_H^2}
\end{equation}
where $v$ is the Higgs field vacuum expectation value.

\end{itemize}

The effective upper bound in energies of the Majorana neutrinos in the galactic centre and haloes cannot exceed the MeV scale (usually is at the keV. Hence, by using as an ultraviolet cutoff $\Lambda \ll M_H \sim O(100) ~{\rm GeV}$, and taking into account that the dominant contributions to the amplitude (\ref{ff}) come from the region of integration $p_i (i=1, \dots 4) \ll k \ll M_H$,
one may estimate the total total four fermion amplitude, stemming from the sum of the terms (\ref{tlec}), (\ref{ff}), as proportional to
\begin{equation}\label{effcoupl}
F_{\alpha 1}^4 \frac{1}{M_H^4} \times \int \frac{d^4 k}{(2\pi)^4} \frac{1}{\slashed{k}} \, \frac{1}{\slashed{k}} \propto F_{\alpha 1}^4 \frac{\Lambda^2}{M_H^4}
\end{equation}
which yields the effective four fermion coupling $G_{4f}$ for this induced interaction.
With the value (\ref{f1v}) of the Yukawa coupling $F_{\alpha 1}$, and a cut-off $\Lambda$  in the ball park of a few MeV, the effective coupling $G_{4f}$ is much smaller than the postulated strong four fermion coupling in our self-interacting model, to be discussed below\footnote{Actually, in view of the smallness of $F_{\alpha 1}$, the same conclusion holds for $\Lambda$ even much larger than the electroweak scale $M_W =O(100)$~GeV.}. Hence the conventional $\nu$MSM model, without the inclusion of relatively strong self interactions, as proposed here, cannot provide an effective description of the galactic structure.


\end{document}